\documentclass[12pt,a4paper]{article}
\date{\today}

\newcommand{\insertplot}[5]{\begin{figure}
 \hfill\hbox to 0.05in{\vbox to #5in{\vfill
 \inputplot{#1}{#4}{#5}}\hfill}
 \hfill\vspace{-.1in}
 \caption{#2}\label{#3}
 \end{figure}}
 \newcommand{\inputplot}[3]{
 \special{ps: plotfile #1}
}
\newcounter{fig}   \newcommand{\lbfig}[1]{\refstepcounter{fig}
\label{#1} }

\usepackage{epsfig}
\usepackage{amsmath}
\usepackage{amsfonts}
\usepackage{graphicx}
\usepackage[german, english]{babel}
\usepackage{a4wide}
\usepackage{amsmath}
\usepackage{amssymb}
\usepackage{ifthen}
\usepackage{epsfig}

\begin{document}

\newcommand{\dd}{\mbox{d}}
\newcommand{\tr}{\mbox{tr}}
\newcommand{\la}{\lambda}
\newcommand{\ta}{\theta}
\newcommand{\f}{\phi}
\newcommand{\vf}{\varphi}
\newcommand{\ka}{\kappa}
\newcommand{\al}{\alpha}
\newcommand{\ga}{\gamma}
\newcommand{\de}{\delta}
\newcommand{\si}{\sigma}
\newcommand{\bomega}{\mbox{\boldmath $\omega$}}
\newcommand{\bsi}{\mbox{\boldmath $\sigma$}}
\newcommand{\bchi}{\mbox{\boldmath $\chi$}}
\newcommand{\bal}{\mbox{\boldmath $\alpha$}}
\newcommand{\bpsi}{\mbox{\boldmath $\psi$}}
\newcommand{\brho}{\mbox{\boldmath $\varrho$}}
\newcommand{\beps}{\mbox{\boldmath $\varepsilon$}}
\newcommand{\bxi}{\mbox{\boldmath $\xi$}}
\newcommand{\bbeta}{\mbox{\boldmath $\beta$}}
\newcommand{\ee}{\end{equation}}
\newcommand{\eea}{\end{eqnarray}}
\newcommand{\be}{\begin{equation}}
\newcommand{\bea}{\begin{eqnarray}}
\newcommand{\ii}{\mbox{i}}
\newcommand{\e}{\mbox{e}}
\newcommand{\pa}{\partial}
\newcommand{\Om}{\Omega}
\newcommand{\vep}{\varepsilon}
\newcommand{\bfph}{{\bf \phi}}
\newcommand{\lm}{\lambda}
\def\theequation{\arabic{equation}}
\renewcommand{\thefootnote}{\fnsymbol{footnote}}
\newcommand{\re}[1]{(\ref{#1})}
\newcommand{\R}{{\rm I \hspace{-0.52ex} R}}
\newcommand{\N}{{\sf N\hspace*{-1.0ex}\rule{0.15ex}%
{1.3ex}\hspace*{1.0ex}}}
\newcommand{\Q}{{\sf Q\hspace*{-1.1ex}\rule{0.15ex}%
{1.5ex}\hspace*{1.1ex}}}
\newcommand{\C}{{\sf C\hspace*{-0.9ex}\rule{0.15ex}%
{1.3ex}\hspace*{0.9ex}}}
\newcommand{\eins}{1\hspace{-0.56ex}{\rm I}}
\renewcommand{\thefootnote}{\arabic{footnote}}

\title{
\begin{flushright}\ \vskip -2cm {\small {\em DCPT-11/03}}\end{flushright}
\vskip 2cm Q-vortices, Q-walls and coupled Q-balls}
\author{
Ya Shnir$^{\dagger\star}$\\[10pt]
$^\dagger$ {\small Institut f\"ur Physik, Universit\"at Oldenburg, Germany}
\\ $^{\star}$ {\small Department of Mathematical Sciences, Durham University, UK}}
\date{January 2011}
\maketitle

\begin{abstract}
{\sf We discuss three different globally regular
non-topological stationary soliton solutions in the theory of a complex scalar field  in 3+1 dimensions,
so-called Q-balls, Q-vortices and Q-walls. The
charge, energy and profiles of the corresponding solutions
are presented for each configuration studied.
The numerical investigation of these three types of solutions shows different behavior of
charge and energy with changing frequency $\omega$  for each type.
We investigate properties of new families of coupled non-topological 2-Q-ball solutions obtained within
the same model by generalization of the ansatz for the scalar field
which includes an independent phase.
New composite solutions for another known model, which describes two Q-balls
minimally interacting via a coupling term, are discussed briefly.}
\end{abstract}


\section{Introduction}
Q-balls are stationary localized non-topological soliton solutions
of a nonlinear field theory which carry global $U(1)$ charge \cite{Coleman:1985ki}.
In order to evade Derrick's theorem \cite{Derrick}, Q-ball configurations must be time-dependent.
In the simplest case there is a single complex self-interacting scalar field
with an explicitly time-dependent phase and sextic non-renormalizable potential.
The charge then is directly proportional to the frequency of rotation.

As shown in \cite{Volkov:2002aj}, it is possible to construct spinning axially symmetric generalization
of the Q-ball solutions. These solutions correspond to stationary localized configurations
possessing a finite mass and a finite angular momentum which is
quantized, $J = nQ$, where $Q$ is the Noether charge of the solution and $n \in \mathbb{Z}$
corresponds to the winding around the symmetry axis.
Possessing even or odd parity, their energy density forms one or more tori \cite{Radu:2008pp}.
Recently the properties of these rotating solutions in the presence of gravity were considered in
\cite{KKL1,Kleihaus:2007vk}.

An interesting class of non-topological solitons having axial symmetry was discussed in
\cite{Volkov:2002aj}. These solutions
represent stationary Q-vortices. Other configurations which possess planar symmetry
has been investigated in \cite{MacKenzie:2001av}. In the latter case the solutions are refer to as
Q-walls.

The Q-ball solutions arise in various models, however one of the most interesting
examples is related with the supersymmetric extensions of the standard model
with flat directions in their scalar potentials \cite{Kusenko:1997zq}. In such a case
the $U(1)$ charge is associated with the symmetries of baryon and lepton number conservation, so
the corresponding solutions are leptonic and baryonic balls. It was suggested \cite{Kusenko:1997si}
that the Q-balls arising in a supersymmetric model may play a role in baryogenesis
through the Affleck-Dine baryogenesis mechanism \cite{Affleck}.

Since the supersymmetric models contain several scalar fields, they interact via various potentials.
This pattern can then lead to the complicated picture of evolution of the Q-balls. On the other hand,
the system of coupled Q-balls may possess new solutions whose properties may be rather exotic.
For example previous research has suggested that such a system may support
the existence of `twisted' Q-balls which would have a certain similarity
with twisted loops in the Faddeev-Skyrme model \cite{Radu:2008pp}. An intriguing observation is that
in the case of the supersymmetric models the flat Q-ball potential yields
the energy/charge relation  $E \sim Q^{3/4}$ \cite{Kusenko:1997zq,Dvali:1998},
which precisely matches the topological energy bound
for the solitons of the Faddeev-Skyrme model \cite{Vakulenko}.
Another interesting similarity between these models is that the spectrum of solition solutions in both cases
includes not only the fundamental localised solitons, the Q-ball and the hopfions, respectively, but also
extended objects like vortices \cite{Volkov:2002aj,Hietarinta:2003vn} and walls \cite{MacKenzie:2001av,Ward}.

In this paper some properties of Q-ball type solitons are studied.
First, we here consider properties of the corresponding solutions to the same
simple model with non-renormalisable $|\Phi|^6$-potential.
We discuss both the usual spherically symmetric Q-balls
and extended objects, Q-vortices  and Q-walls.
Another purpose of this paper is to investigate properties of the `twisted' Q-balls using the parametrization
suggested in \cite{Radu:2008pp}. We argue that these solutions may exist only if
the constituents are actually two copies of the same configuration.
Using another model suggested by Brihaye and Hartmann \cite{Brihaye:2007tn}
we present numerical arguments for the existence of a class of
coupled 2-Q-ball configurations which may have different geometry. We present both spinning
and non-spinning solutions and briefly discuss their properties.

The plan of the paper is as follows. In section II, we review the model and give the equations
and boundary conditions. Here we discuss the Q-ball, Q-vortex and Q-wall solutions for the
model with sextic potential, while in section III, we discuss our results for two different models
with two interacting components. We present our conclusions in section IV.

\section{The model and fundamental solutions}
We consider a theory of a complex scalar field $\Phi$ in (3+1)-dimensional flat space–time defined
by the Lagrangian density
\be \label{model}
L = \partial_\mu \Phi \partial^\mu \Phi^* - U(|\Phi|)
\ee
with a $U(1)$ invariant scalar potential $U(|\Phi|)$.

To secure the existence of non-topological solitons, the vacuum of the model should be
non-degenerate \cite{Lee:1991ax}. In particular, a potential polynomial in  $|\Phi|^2$
should contain powers of $\Phi$ higher than four \cite{Coleman:1985ki}, which makes the model non-renormalizable.
We will consider the $|\Phi|^6$ potential \cite{Volkov:2002aj,Radu:2008pp,Multamaki:1999an}
\be \label{pot}
U(|\Phi|) =  a|\Phi|^6 + b|\Phi|^4+c |\Phi|^2,
\ee
where $a=1, b=-2, c=1.1$, i.e. $U(|\Phi|) = |\Phi|^2[(1-|\Phi|^2)^2 +0.1]$.
This is a typical potential in field theories that can contain extended objects of the Q-ball type.
So, the mass of the scalar excitation is $m^2= \frac{1}{2}U''(0)= c$.

The corresponding field equation is
\be \label{equation}
\partial_\mu  \partial^\mu \Phi  + \frac{ \partial U}{\partial \Phi^*} = 0,
\ee
and the global $U(1)$ Noether charge is
\be \label{charge}
Q = i \int d^3 x (\Phi \dot \Phi^* - \Phi^* \dot \Phi)
\ee

Depending on the boundary conditions, different types of solutions can be found.
The well known Q-ball solution corresponds to a minimum of the energy functional at a fixed
charge \cite{Coleman:1985ki}.
Assuming that $\Phi$ depends on time harmonically
\be \label{ansatz}
\Phi = \phi({\bf r}) e^{i\omega t} \, ,
\ee
where $\omega$ is the internal rotation frequency and $\phi({\bf r})$ is time independent real function of
coordinates ${\bf r}$, we get the total energy functional
\be \label{energy}
T_0^0= E = \int d^3x \left( \omega^2 \phi^2 + (\nabla \phi)^2 + U(\phi) \right) \, ,
\ee
which effectively describes the motion in the potential $V(\phi) =  \omega^2 - U(\phi)$.
The usual restriction on the rotation frequency is that the potential $U(\phi)$ should have its
absolute minimum $U(0)=0$
and the Q-ball solution must decay at spatial infinity. This implies \cite{Radu:2008pp}
\be \label{omega}
\omega_-^2 < \omega^2 < \omega_+^2 \, ,
\ee
where with our choice of the parameters, $\omega_-^2 = 0.1$ and $\omega_+^2 = m^2 =1.1$.

The simplest spherically symmetric Q-ball placed at the origin is a stationary solution of the form
$\Phi = f(r)e^{i\omega t}$ \cite{Coleman:1985ki},
where the real amplitude $f(r)$ satisfies the equation
\be
\frac{d^2 f}{dr^2} + \frac{2}{r} \frac{df}{dr} + \omega^2 f = \frac{1}{2} \frac{\partial U}{\partial f}
\, .
\ee
The regular solution of this equation asymptotically decays as $f(r) \sim \frac{1}{r}e^{-\sqrt{\omega_+^2 - \omega^2}}$.
The charge of the fundamental Q-ball depends on $\omega$ as
\be
Q(\omega)=8\pi \omega \int\limits_0^\infty dr r^2 \phi^2\, .
\ee
As $\omega^2 \to \omega_-^2$  the Q-ball expands and its charge diverges. In the opposite limit $\omega^2 \to \omega_+^2 = m^2$
the Q-ball also expands. The maximal value of the function $f(0)$ at the origin
then tends to zero although the integrated
charge/energy density diverges (See Fig.~\ref{fig:1}~(a)).

Another stationary solution of the model \re{model} is the vortex configuration discussed in \cite{Volkov:2002aj}.
Assuming the field has cylindrical symmetry, this leads us to the parametrization
\be \label{ansatz-vort}
\Phi = g(\rho)e^{i\omega t}, ~{\rm where}~~ \rho = \sqrt{x^2+y^2}
\ee
Then the field equation becomes
\be \label{equation-rho}
\frac{d^2 g}{d\rho^2} + \frac{1}{\rho} \frac{d g}{d\rho} + \omega^2 g = \frac{1}{2} \frac{\partial U}{\partial g}
\ee
with the boundary conditions $g'(0)=0$ and $g(\infty)=0$. So, the field $g(\rho)$ has a maximum at $\rho = 0$
and decreases monotonically towards zero at infinity.
\begin{figure}[hbt]
\lbfig{fig:1}
\begin{center}
(a)\hspace{-0.6cm}
\includegraphics[height=.22\textheight, angle =-90]{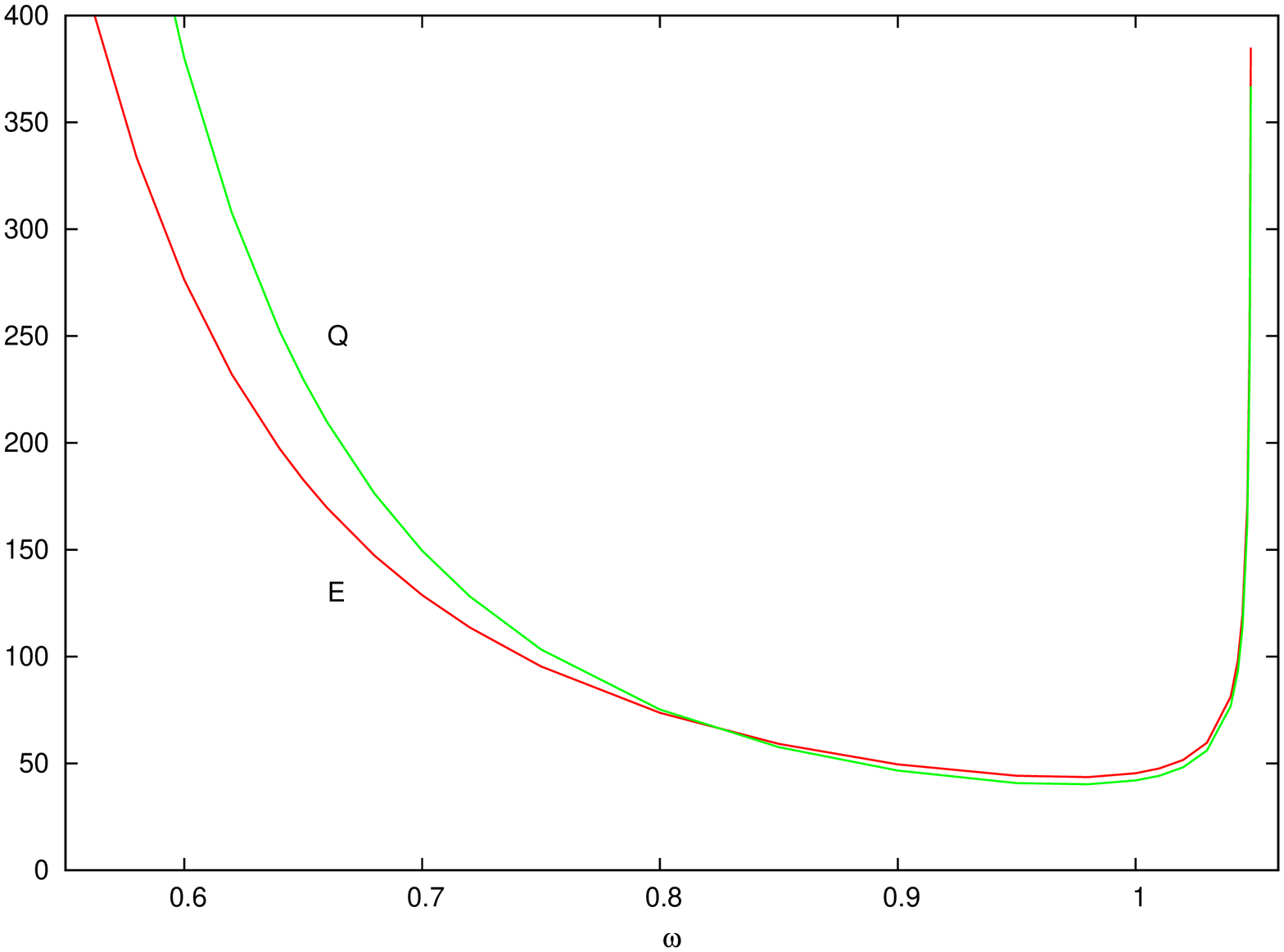}
\hspace{0.5cm} (b)\hspace{-0.6cm}
\includegraphics[height=.22\textheight, angle =-90]{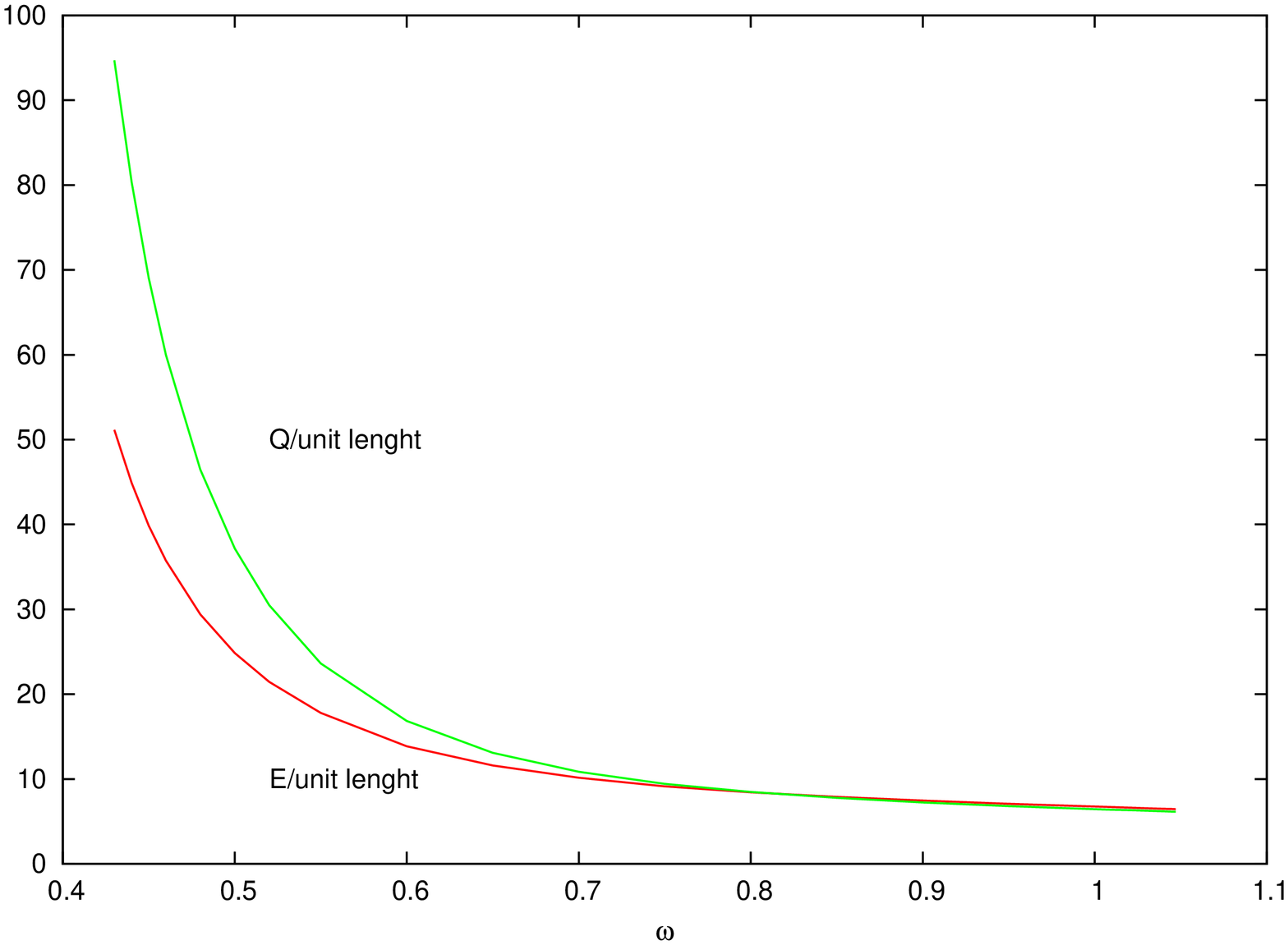}
\hspace{0.5cm} (c)\hspace{-0.6cm}
\includegraphics[height=.22\textheight, angle =-90]{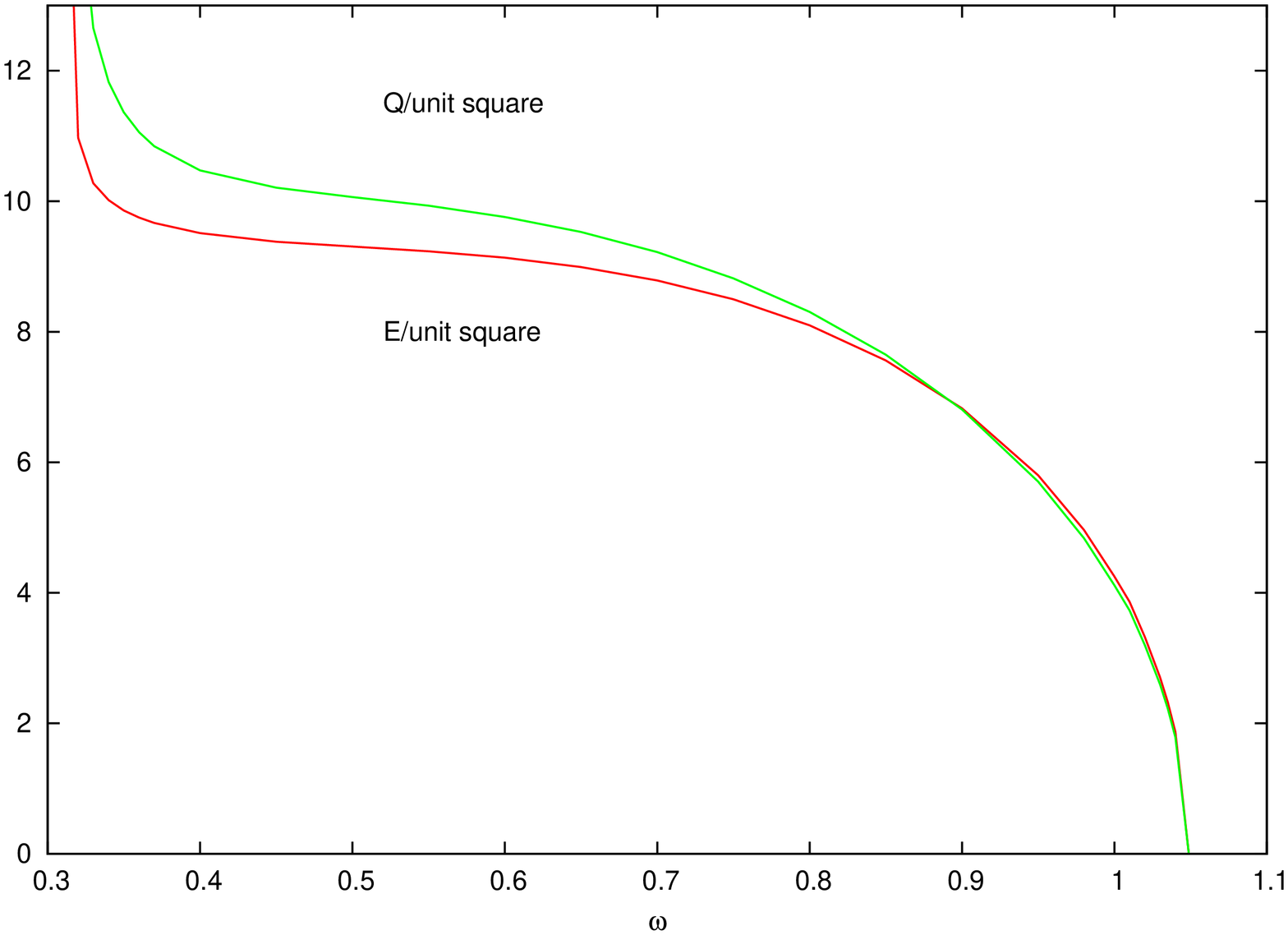}
\end{center}
\vspace{-0.5cm}
\caption{\small Stationary non-spinning non-topological solitons. The energy $E$ and the charge $Q$ are shown
as functions of the frequency $\omega$ for the spherically symmetric Q-ball (a); Q-vortex (per unit length) (b); and
Q-wall (per unit area) (c).
}
\end{figure}

The regular vortex solution exist for a frequency within the interval \re{omega}. It has properties similar to
the spherically symmetric Q-ball, in particular as $\omega^2 \to \omega_-^2$  the Q-vortex expands, its charge and its energy
diverge. However, it is plausible that
before approaching this limit the vortex may become unstable with respect to linear perturbations, it could decay into
Q-balls whose energy per unit length is smaller. In the opposite limit $\omega^2 \to \omega_+^2$ the vortex also expands,
however, as seen in Fig.\ref{fig:3}, in this limit the value of the field at
the origin $\Phi(0)$ decreases and both the
integrated energy per unit length
\be
E=2\pi \int\limits_0^\infty d\rho~ \rho\left[ (g')^2 + \omega^2 g^2 + U(g) \right]
\ee
and the integrated charge per unit length approach constant
values, as illustrated in Fig.\ref{fig:1}~(b).

Another configuration can be constructed by imposing condition of planar invariance \cite{MacKenzie:2001av}.
Then the field $\Phi$ does not depend on $\rho$ and we can use the ansatz
\be \label{ansatz-wall}
\Phi = h(z)e^{i\omega t}\, .
\ee
Substituting this ansatz into the field equation we obtain the equation without the `friction' term
\be \label{equation-z}
\frac{d^2 h}{dz^2} + \omega^2 h = \frac{1}{2} \frac{\partial U}{\partial h}
\ee
which corresponds to the globally regular `Q-wall' solution discussed in a different model in \cite{MacKenzie:2001av}.
\begin{figure}[hbt]
\lbfig{fig:6}
\begin{center}
(a)\hspace{-0.6cm}
\includegraphics[height=.33\textheight, angle =-90]{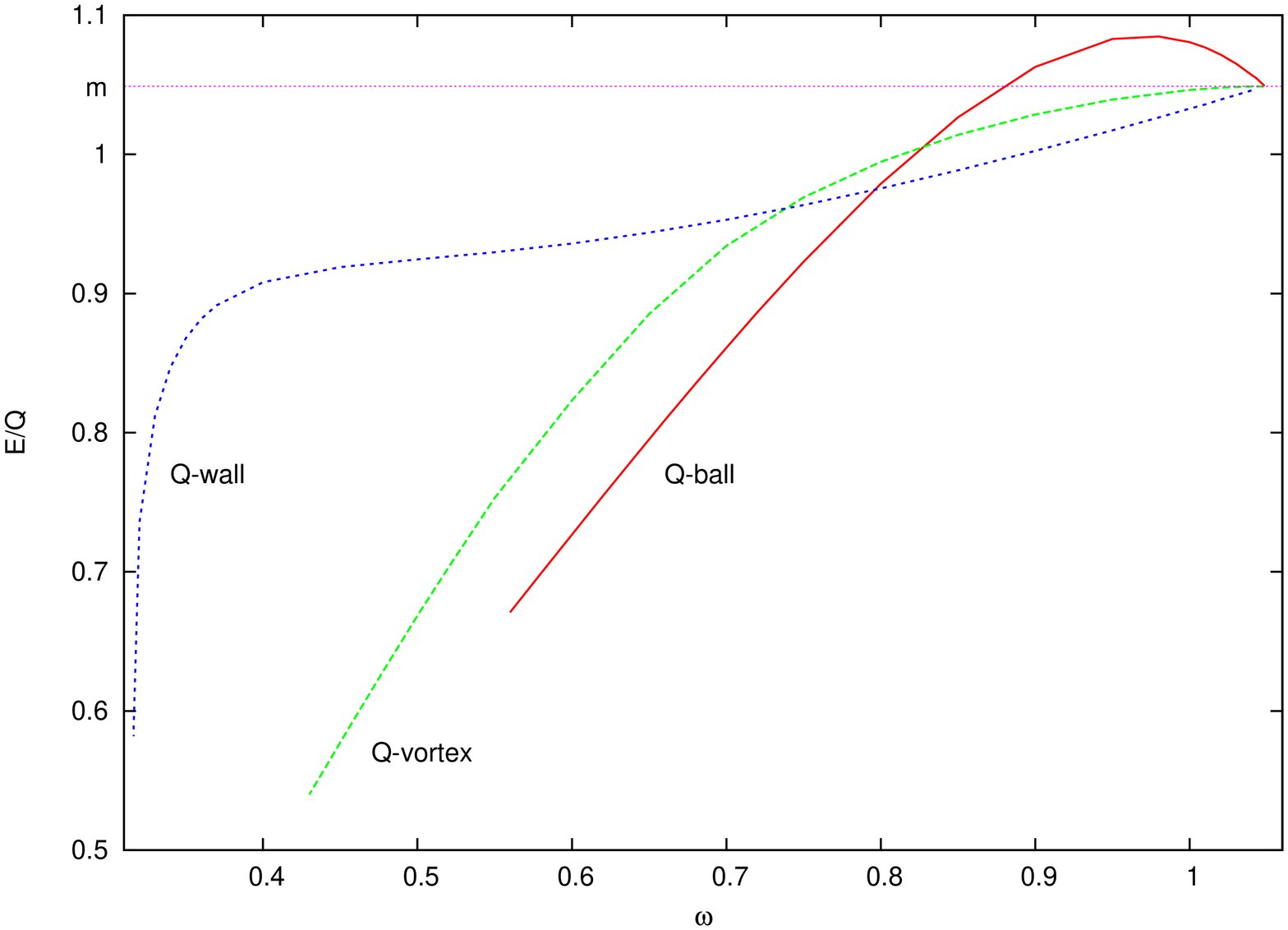}
\hspace{0.5cm} (b)\hspace{-0.6cm}
\includegraphics[height=.33\textheight, angle =-90]{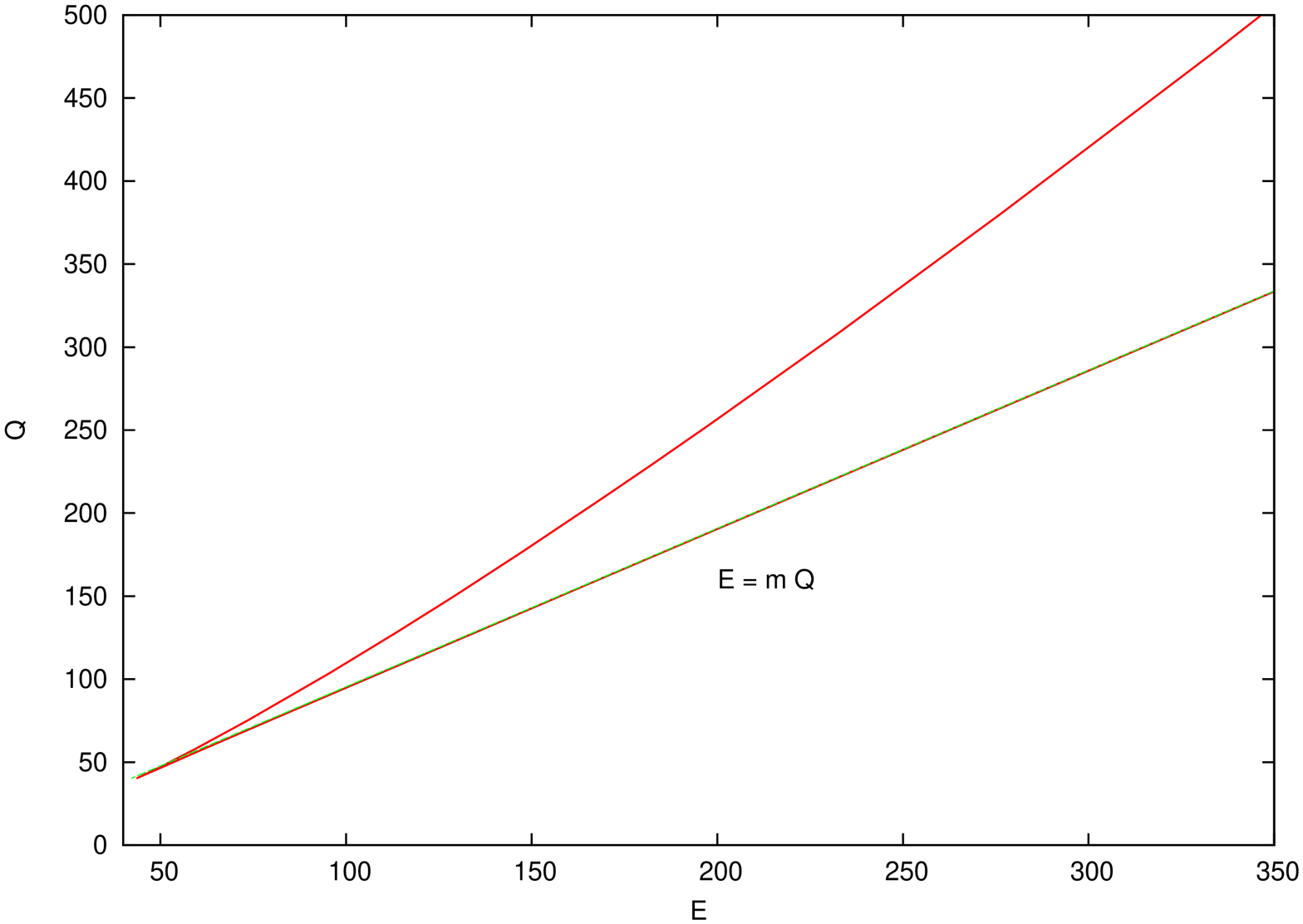}
\hspace{0.5cm} (c)\hspace{-0.6cm}
\includegraphics[height=.33\textheight, angle =-90]{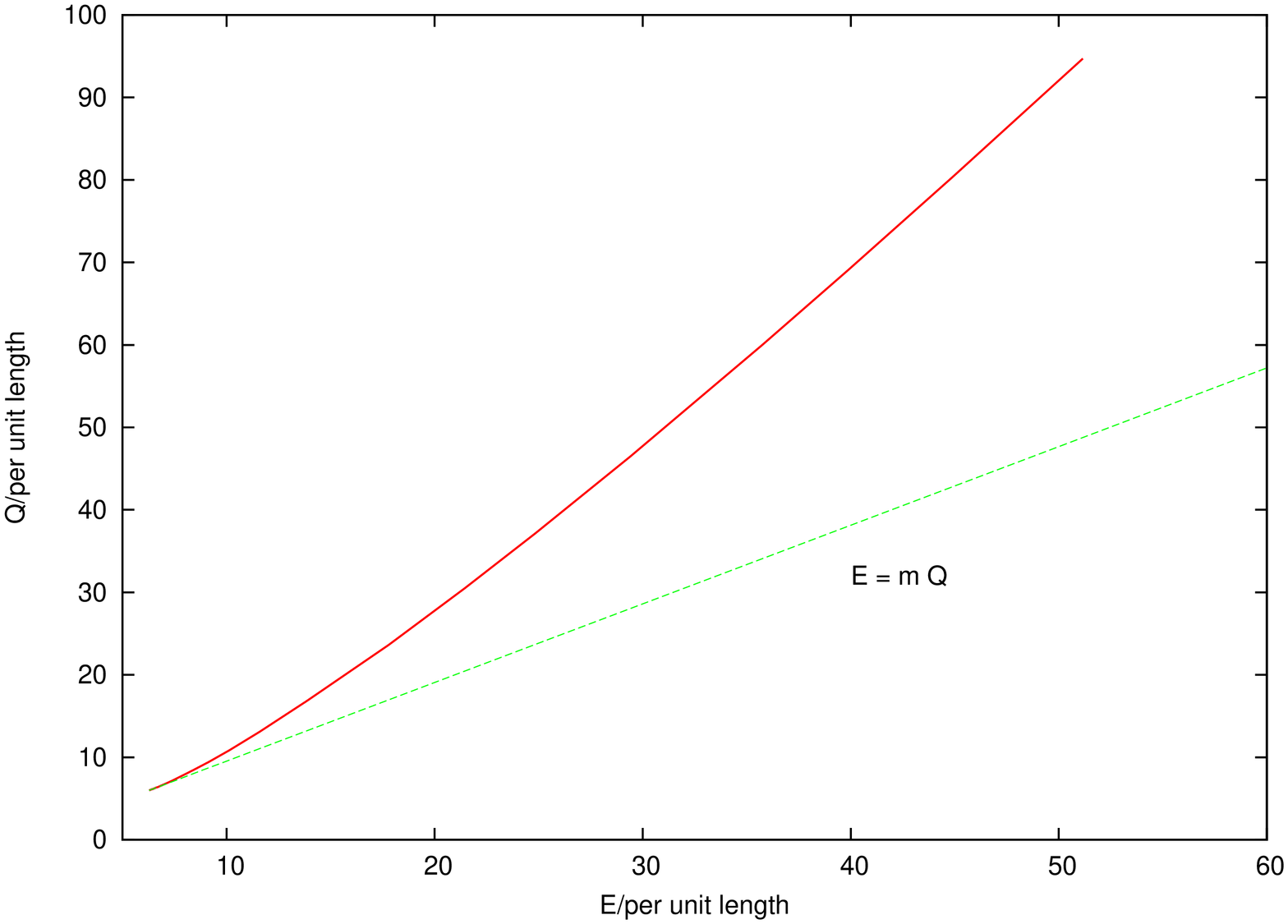}
\hspace{0.5cm} (d)\hspace{-0.6cm}
\includegraphics[height=.33\textheight, angle =-90]{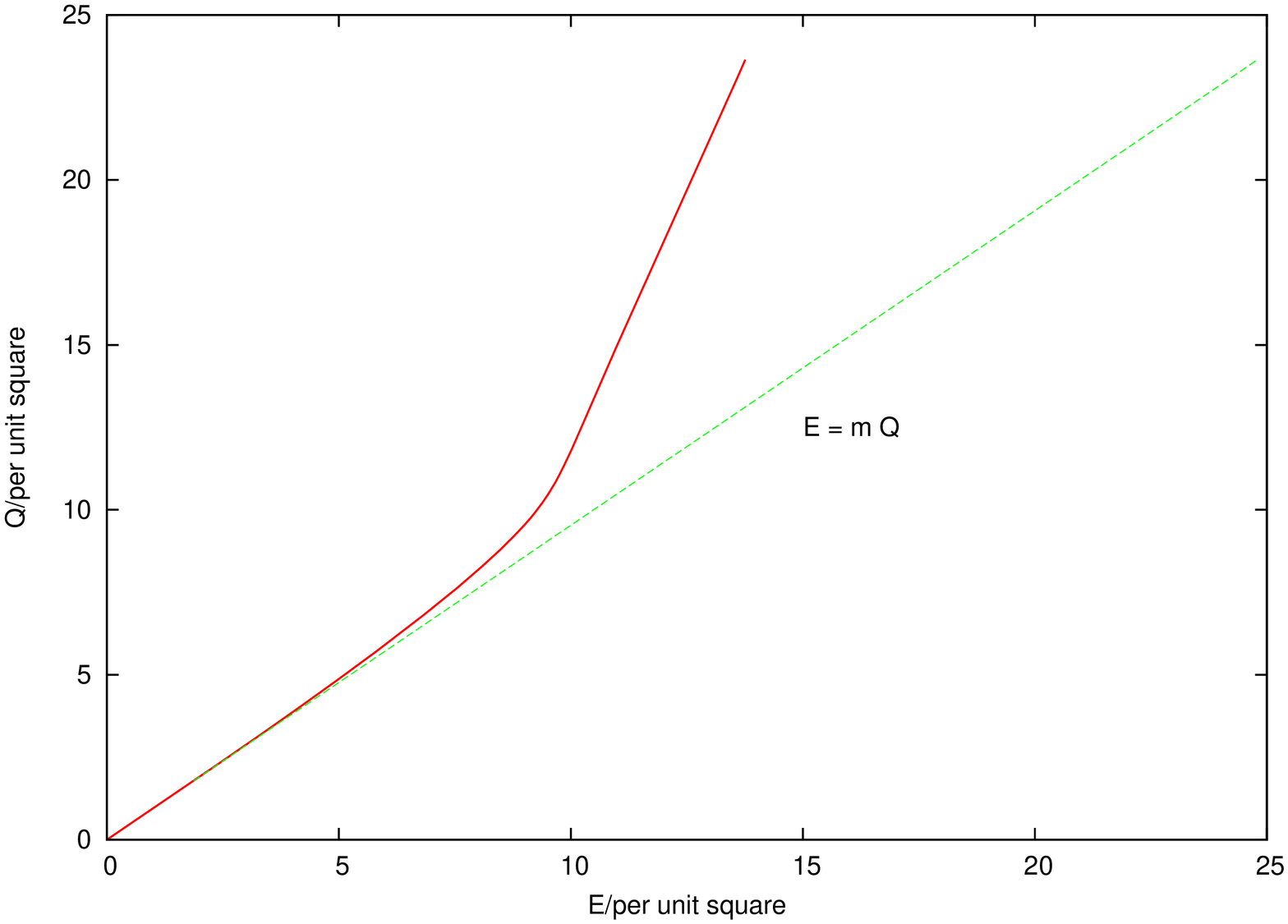}
\end{center}
\vspace{-0.5cm}
\caption{\small Energy-charge ratio versus
$\omega$ (a) and the energy versus charge are shown
for the spherically symmetric Q-ball (b), Q-vortex (per unit length)(c)  and
Q-wall (per unit area) (d). A straight dashed line $E=mQ$ indicating the margin of stability is drawn.
}
\end{figure}

For this configuration the field $h(z)$ smoothly
interpolates between some finite value on the z-axis $h(0)$ and zero value as $z \to \infty$.
Indeed, linearisation of the equation \re{equation-z} around $h=0$ gives asymptotically
$h \sim e^{-\sqrt{(\omega_+^2-\omega^2)} z}$, so the field decays exponentially fast. The solution also is
regular at $z=0$ since $h'(0)=0$.

As seen in Fig.\ref{fig:3},
the value of $h(0)$ depends on $\omega$; as $\omega^2 \to \omega_+^2$ it smoothly decreases towards zero although the
Q-wall expands. In this limit the energy and the charge of the Q-wall per unit area tends to zero,
as illustrated in Fig \ref{fig:1}~(c). As $\omega^2$ starts to decrease
the energy density of the wall per unit area increases forming a single maximum at the origin.

\begin{figure}[hbt]
\lbfig{fig:2}
\begin{center}
\hspace{0.5cm} (a) \hspace{6.0cm} (b) \hspace{6.5cm} (c) \\
\vspace{0.5cm}
\includegraphics[height=.22\textheight, angle =0]{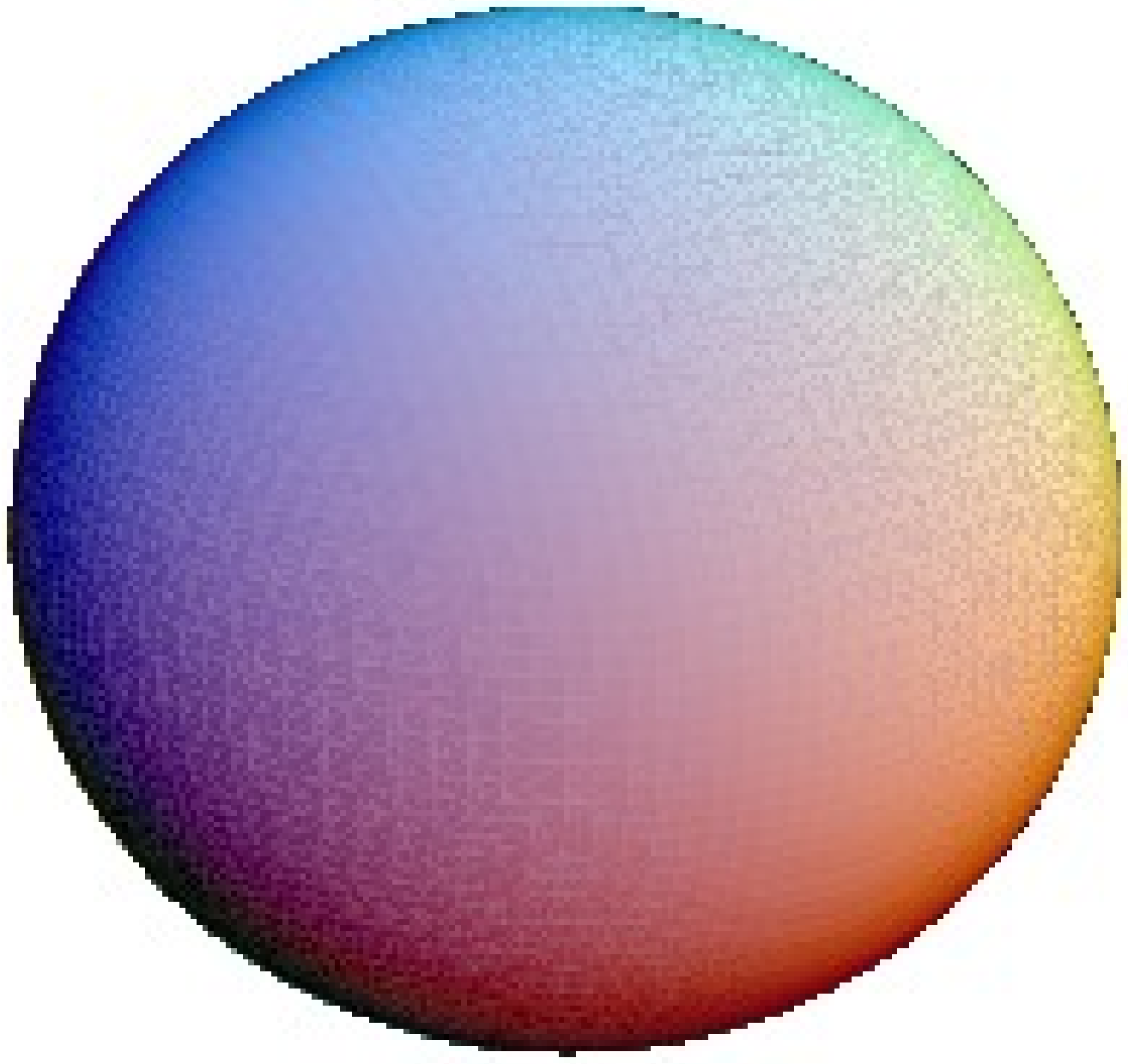}
\includegraphics[height=.22\textheight, angle =0]{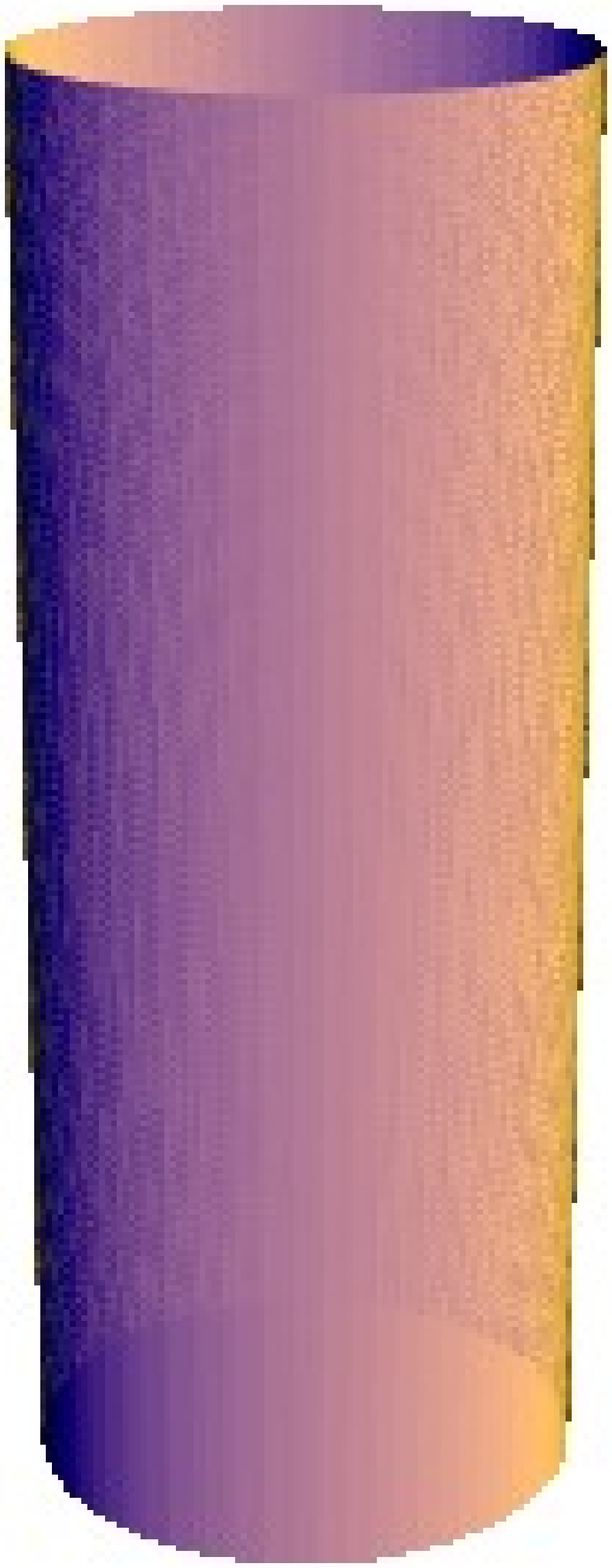}
\includegraphics[height=.20\textheight, angle =0]{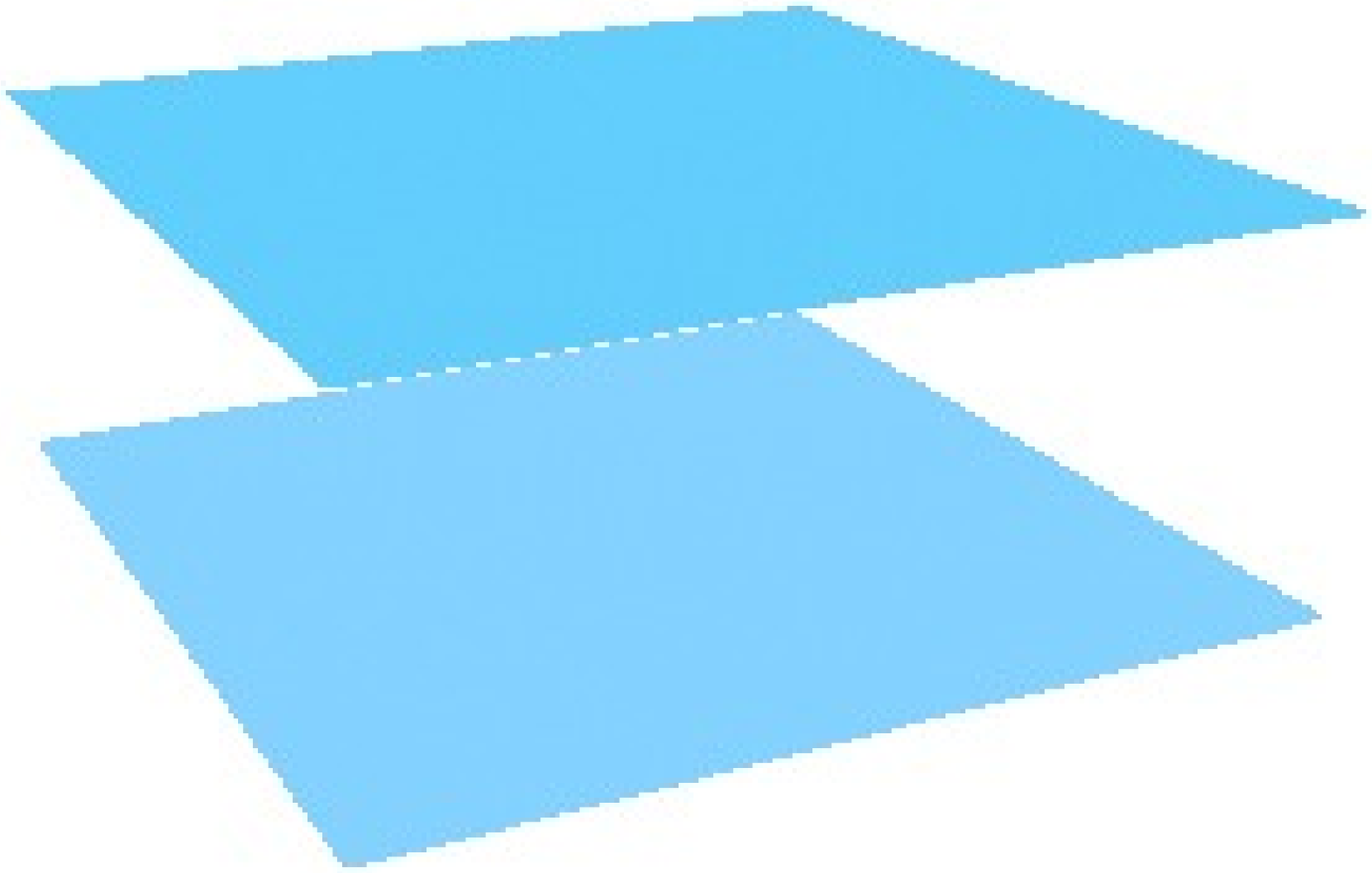}
\end{center}
\caption{\small The energy isosurfaces  of the stationary non-spinning Q-ball (a), Q-vortex (b) and Q-wall (c) are shown for
$\omega=0.8$.
}
\end{figure}

In the opposite limit $\omega^2 \to \omega_-^2$ the wall rapidly expands forming
new vacuum $\Phi_{vac} = 1 > \Phi(\infty)$ around the origin, as seen from Figures \ref{fig:4}, \ref{fig:5}.
Then both the charge $Q$ and the energy $E$ per unit area are growing without limit (Fig.~\ref{fig:1} (c)), so
in all the cases there is a different behavior as $\omega^2$ tends to $\omega_+^2 = m^2$.

The fundamental solutions of the model \re{model} are stable under condition $E< mQ$ \cite{Coleman:1985ki}.
In Fig.~\ref{fig:6} we have plotted the energy $E$ as a function of $Q$ and the ratio $E/Q$ versus $\omega$ for the different
types solutions. Here we have also plotted a straight line $E=mQ$ indicating the region of stability of the configurations with respect
to decays into quanta of the scalar field, so the fundamental $Q$-ball becomes unstable as
$\omega$ approaches the critical value
$\omega_{cr} \approx 0.89$. Charge and energy then approach a minimum forming a typical spike profile
(see Fig.\ref{fig:6}, frame (b)) which indicates the saddle node bifurcations and instability of the upper branch. On the other hand,
both the Q-vortices and Q-walls remain stable over most of the range of values of $\omega$ (see Fig.~\ref{fig:6},
frames (c) and (d)).
Note also a peculiar kink which appears on the upper branch of the frame (d) representing the Q-wall energy per unit
area versus charge per unit area.
\begin{figure}[hbt]
\lbfig{fig:3}
\begin{center}
\includegraphics[height=.38\textheight, angle =-90]{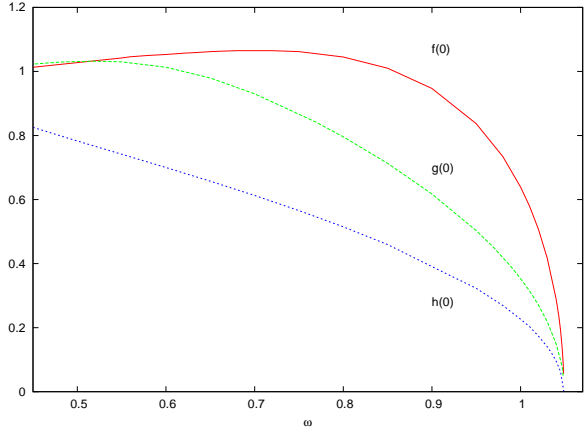}
\end{center}
\vspace{-0.5cm}
\caption{\small The  value at the origin of the profile functions
$f(r)$, $g(\rho)$ and $h(z)$ of the fundamental Q-ball, Q-vortex and Q-wall solutions, respectively are shown as a function
of $\omega$.
}
\end{figure}

\begin{figure}[hbt]
\lbfig{fig:4}
\begin{center}
(a)\hspace{-0.6cm}
\includegraphics[height=.32\textheight, angle =-90]{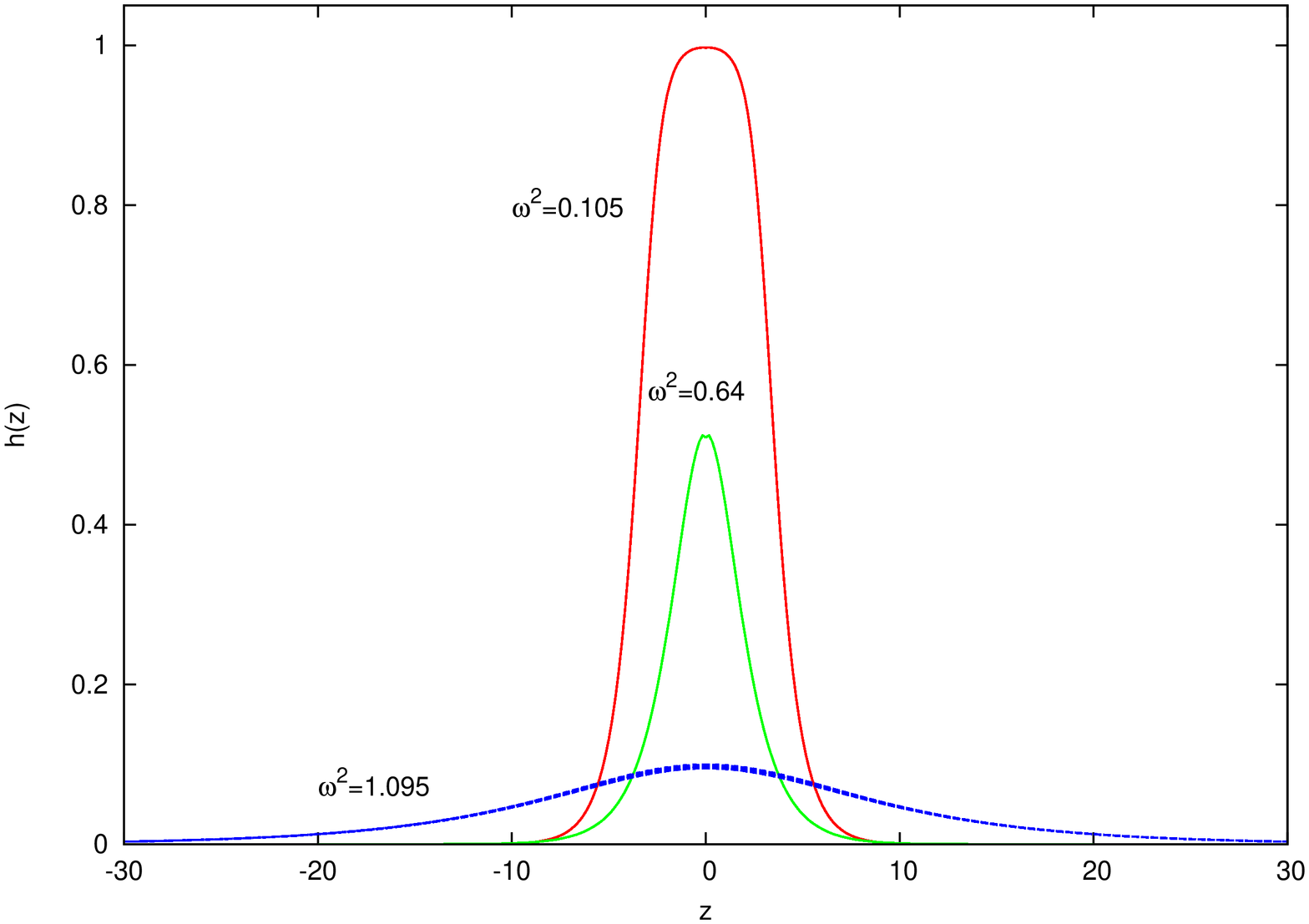}
\hspace{0.5cm} (b)\hspace{-0.6cm}
\includegraphics[height=.32\textheight, angle =-90]{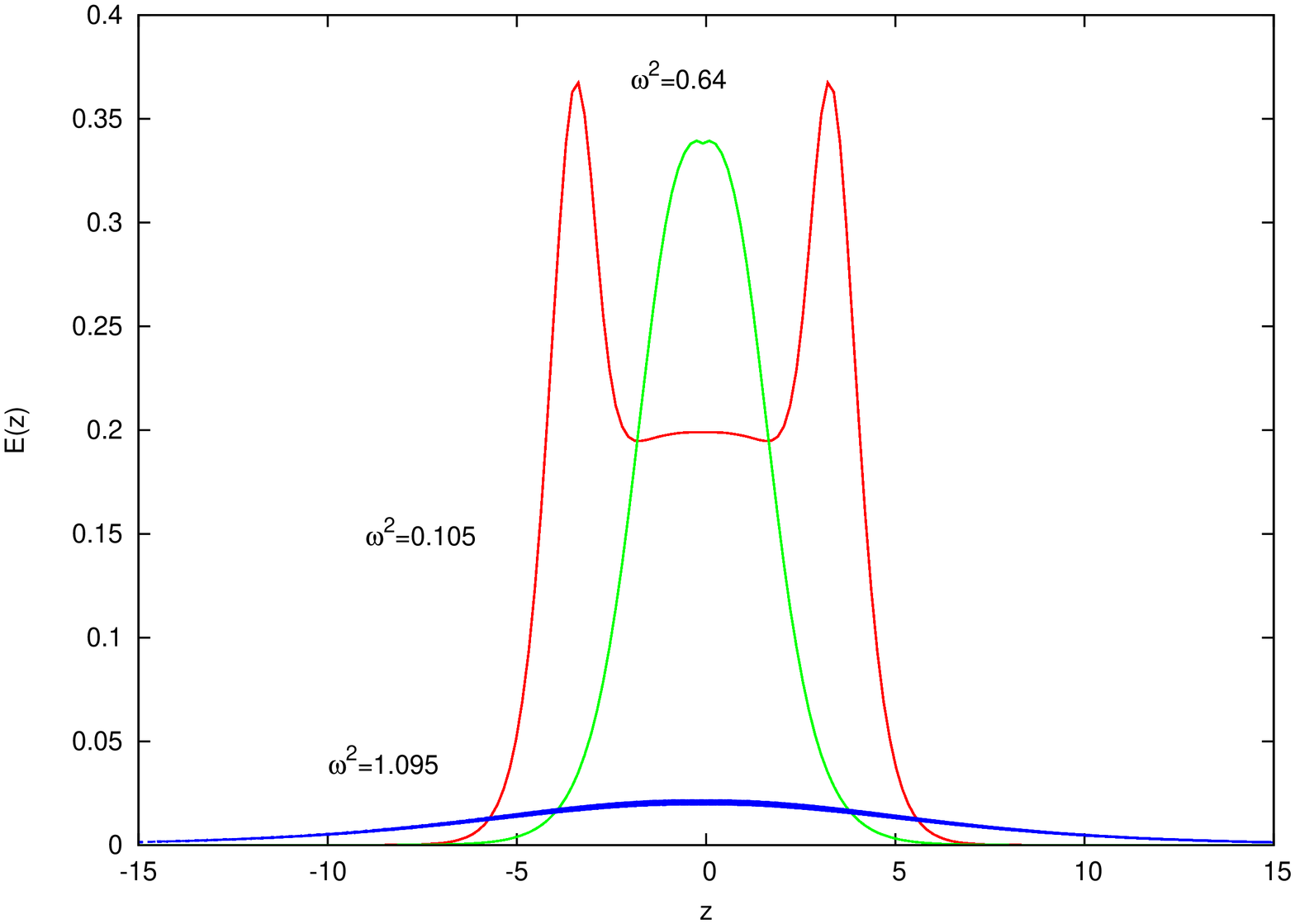}
\end{center}
\vspace{-0.5cm}
\caption{\small The profile of $h(z)$ (a) and the energy density distribution (b)
for the Q-wall solutions is  presented   for $\omega^2 = 1.095, 0.64, 0.105$.}
\end{figure}

\begin{figure}[hbt]
\lbfig{fig:5}
\begin{center}
(a)\hspace{-0.6cm}
\includegraphics[height=.24\textheight, angle =0]{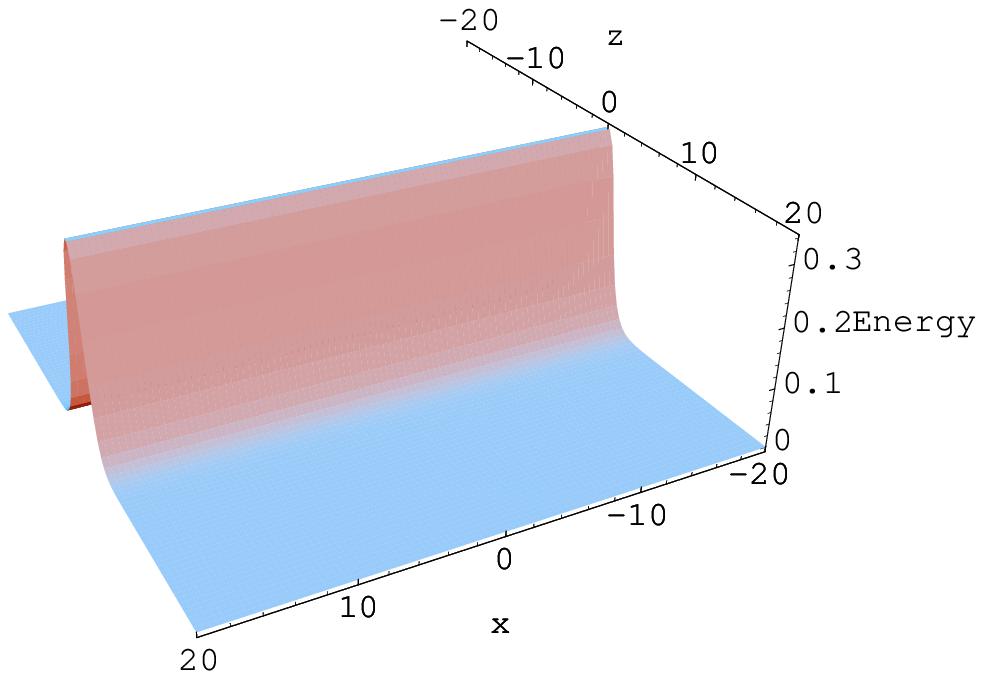}
\hspace{0.5cm} (b)\hspace{-0.6cm}
\includegraphics[height=.24\textheight, angle =0]{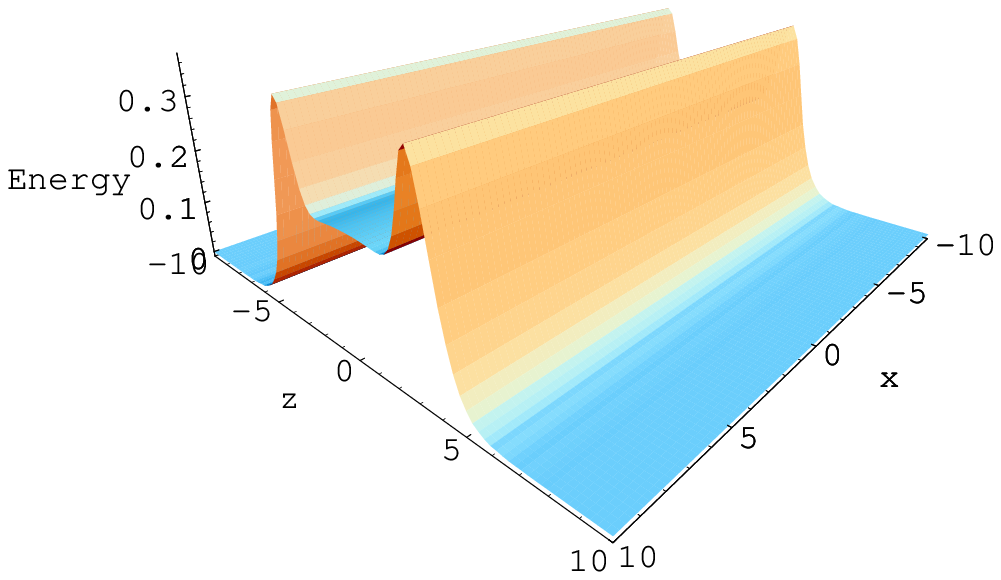}
\end{center}
\vspace{-0.5cm}
\caption{\small The energy density of the Q-wall configuration is shown for $\omega^2 = 0.64$ (a)
and $\omega^2 = 0.105$
as function of $x,z$ coordinates.}
\end{figure}

\section{Coupled Q-balls}
Clearly the relation between the energy and charge of a stationary non-topological soliton of Q-ball type depends
on the explicit form of the potential. In the case of a
flat potential inspired by supersymmetric models \cite{Kusenko:1997zq,Dvali:1998}
the energy/charge relation is $E \sim Q^{3/4}$, which in 3d precisely matches the topological
lower energy bound for the solitons of the Faddeev-Skyrme model \cite{Vakulenko}. Note that this similarity holds not only
in 3 spatial dimensions because for the $d$-dimensional SUSY Q-balls the energy-charge relation is given by
$E \sim Q^{d/(d+1)}$\cite{Dvali:1998}. On the other hand,
for the generalized Faddeev-Skyrme model with a map of degree $k$
$\Phi : \mathbb{R}^{4k-1} \to S^{2k} \subset \mathbb{R}^{2k+1}$ and
the linking number $Q\in \mathbb{Z} \equiv \pi_{4k-1}(S^{2k})$  the generalized Vakulenko-Kapitanski
bound is $E \geq Q^{1-1/(4k)} = Q^{d/(d+1)} $ \cite{Martin}.
Another interesting similarity between these models is that the spectrum of solition solutions in both cases
includes not only the fundamental localised solitons, the Q-balls and the hopfions, respectively, but also
extended objects like vortices \cite{Hietarinta:2003vn,Volkov:2002aj} and walls \cite{Ward,MacKenzie:2001av}.

Recently Radu and Volkov \cite{Radu:2008pp} noted that there is an interesting possibility to
establish a link between some generalization of the model \re{model} and soliton solutions of the
Faddeev-Skyrme model in 3 spacial dimensions. They suggested to consider the generalized spinning
ansatz for the field $\Phi({\bf r})$ which also includes an independent phase $\Psi(r,\theta)$:
\be  \label{spinning-twisted}
\Phi = f(r,\theta)e^{i\omega t - m\Psi(r,\theta) + n \varphi} \equiv [X(r,\theta)+iY(r,\theta)]e^{i\omega t + n \varphi}, ~~~n,m \in \mathbb{Z}.
\ee
Similar to the parametrization of the first two components of the axially symmetric hopfion field \cite{Glad},
the phase function $\Psi(r,\theta)$ increases by $2\pi$ after one revolution around the closed contour
which consists of the $z$-axis and a semi-circle whose radius expands to infinity. The phase $m\Psi + n\varphi$
then increases by  $2\pi m$ times as one moves along this contour and by $2\pi n$ times as one moves on a
circle around the $z$-axis. Such a `twisted' Q-ball is characterised by 2 integer winding numbers $n$ and $m$, which, however do not
have a topological meaning.

Thus with our choice of the parameters of the potential,
the fields $X(r,\theta)$ and $Y(r,\theta)$ of the twisted Q-ball are coupled via the interaction lagrangian
\be \label{coupling}
L_{int}= 3 X^2 Y^2(X^2 + Y^2 - 4/3)
\ee
and the energy density of the stationary system in cylindrical coordinates read
\be \label{tot-eng}
\begin{split}
E &= (\partial_\rho X)^2 + (\partial_\rho Y)^2 + (\partial_z X)^2 + (\partial_z Y)^2 +
\left( \frac{n^2}{\rho^2} + \omega^2\right)(X^2+Y^2) \nonumber\\
&+ U_1[X] + U_2[Y] + L_{int}\, ,
\end{split}
\ee
where
\be \label{pot-2}
U_1[X] = aX^6 + bX^4+c X^2; \quad U_2[Y] = aY^6 + bY^4+c Y^2 \, .
\ee
Actually we are considering two coupled configurations of the Q-ball type with two similar sextic potentials \re{pot}.
This coupling, however, is rather restrictive, unlike in the previous examples considered in
\cite{Brihaye:2007tn} and in \cite{Radu:2008pp}, the parameters of the potentials \re{pot-2} and the
coupling constant are fixed by the structure of the model. Consequently the parameter $\omega$ and the winding
$n$ are the same in both sectors, so the $U(1)$ charge of the configuration is
\be
Q=2\omega \int d^3x (X^2 + Y^2) \, .
\ee

\subsection{Coupled non-spinning $n=0$ solitons}
The ansatz \re{spinning-twisted} provides a natural way to couple two fundamental non-topological solitons.
First, let us consider
coupling of two fundamental solitons of the Q-ball type provided by the ansatz \re{spinning-twisted}
with winding number $n=0$.
The Euler-Lagrange equations arising then from the variations of \re{model}
with respect to the functions $X(\rho, z)$ and
$Y(\rho,z)$ have been integrated by numerically imposing the  boundary
conditions, which respect finite mass-energy and finite energy density
conditions as well as regularity and symmetry requirements.

The numerical calculations are performed employing the package FIDISOL/CADSOL,
based on the Newton-Raphson iterative procedure \cite{FIDISOL}.
We solve the system of two coupled nonlinear partial differential equations
numerically, on a non-equidistant grid in
$\rho$ and $z$, employing the compact coordinates
$x= \rho/(1+\rho) \in [0:1]$ and $y= z/(1+z) \in [0:1]$. Typical grids used have sizes $85 \times 70$.
The relative errors of the solutions are of order of $10^{-4}$ or smaller.

It turns out that the corresponding non-spinning $n=0$ coupled two-component configurations do not exist
for the constituents of different geometry, e.g., we do not find solutions which would
represent the Q-wall coupled to the Q-vortex/Q-ball. We do however find two-component coupled solutions
of the same type with $X(\rho, z)= Y(\rho, z)$ interacting via the potential term \re{coupling}.
We could expect the solutions may exist only in the parameter range $\omega_-^2 < \omega^2 < \omega_+^2$ \re{omega} where
$\omega_-^2 = 0.1$ and $\omega_+^2 = m^2 =1.1$.

Considering two spherically symmetric coupled Q-balls we found that
the corresponding integrated energy and the $U(1)$ charge $Q$
of the system depend on $\omega$ in almost exactly the same way as in the case of the single component model considered in
the previous section. Furthermore, the energy is equally distributed between two components, so the 2-Q-ball system
remains spherically symmetric although the behavior of the profile function $f(0)$ at the origin is different.
This is illustrated in Fig.~\ref{fig:7}, frame (a), where we exhibit the energy-charge
ratios $E/Q$ of the 2-Q-ball system and the single
spherically symmetric Q-ball as functions of $\omega$. Fig.~\ref{fig:7}, frame (b) displays the
evolution of the profile function $f(0)$ at the origin for these systems.

In the limiting cases $\omega \to \omega_+$ and $\omega \to \omega_-$ behavior of the spherically symmetric 2-Q-ball system
is similar to the case of the single spherically symmetric Q-ball. However the potential of interaction of the components
\re{coupling} has a different structure. For $0.90 < \omega < \omega_+$ it is attractive and it has a minimum at the origin,
as seen in Fig.~\ref{fig:8} (b). The depth of the spherical potential well increases as $\omega$ decreases.
For $\omega < 0.90$ the potential of interaction is repulsive at the region around $r=0$,
it has a minimum at $r=r_0 \ne 0$. Thus the energy density distribution of the 2-Q-ball coupled system is forming
a shell structure (see Fig.~\ref{fig:8} (a) where we presented the energy isosurface at $\omega=0.50$).

\begin{figure}[hbt]
\lbfig{fig:7}
\begin{center}
(a)\hspace{-0.6cm}
\includegraphics[height=.33\textheight, angle =-90]{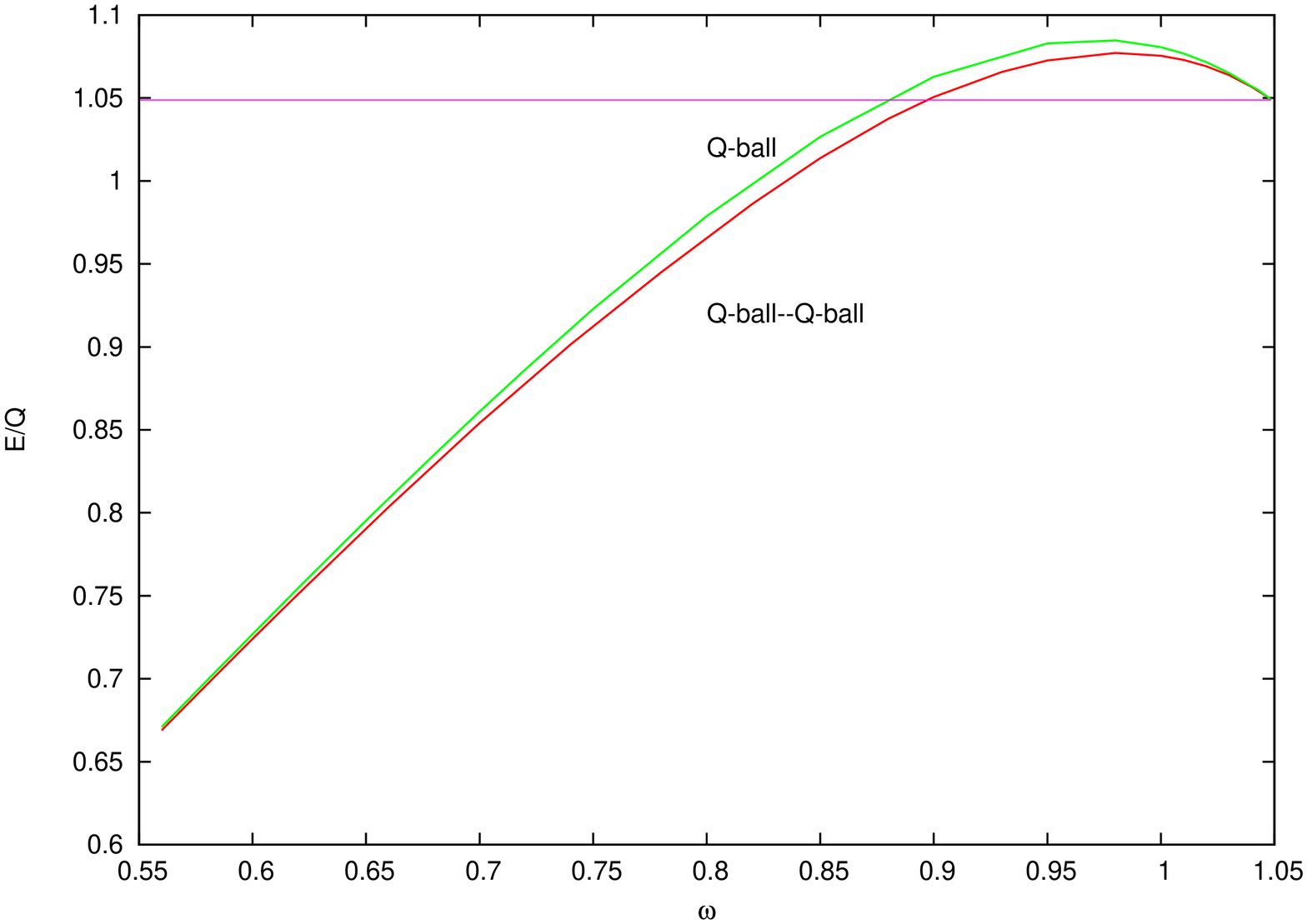}
\hspace{0.5cm} (b)\hspace{-0.6cm}
\includegraphics[height=.33\textheight, angle =-90]{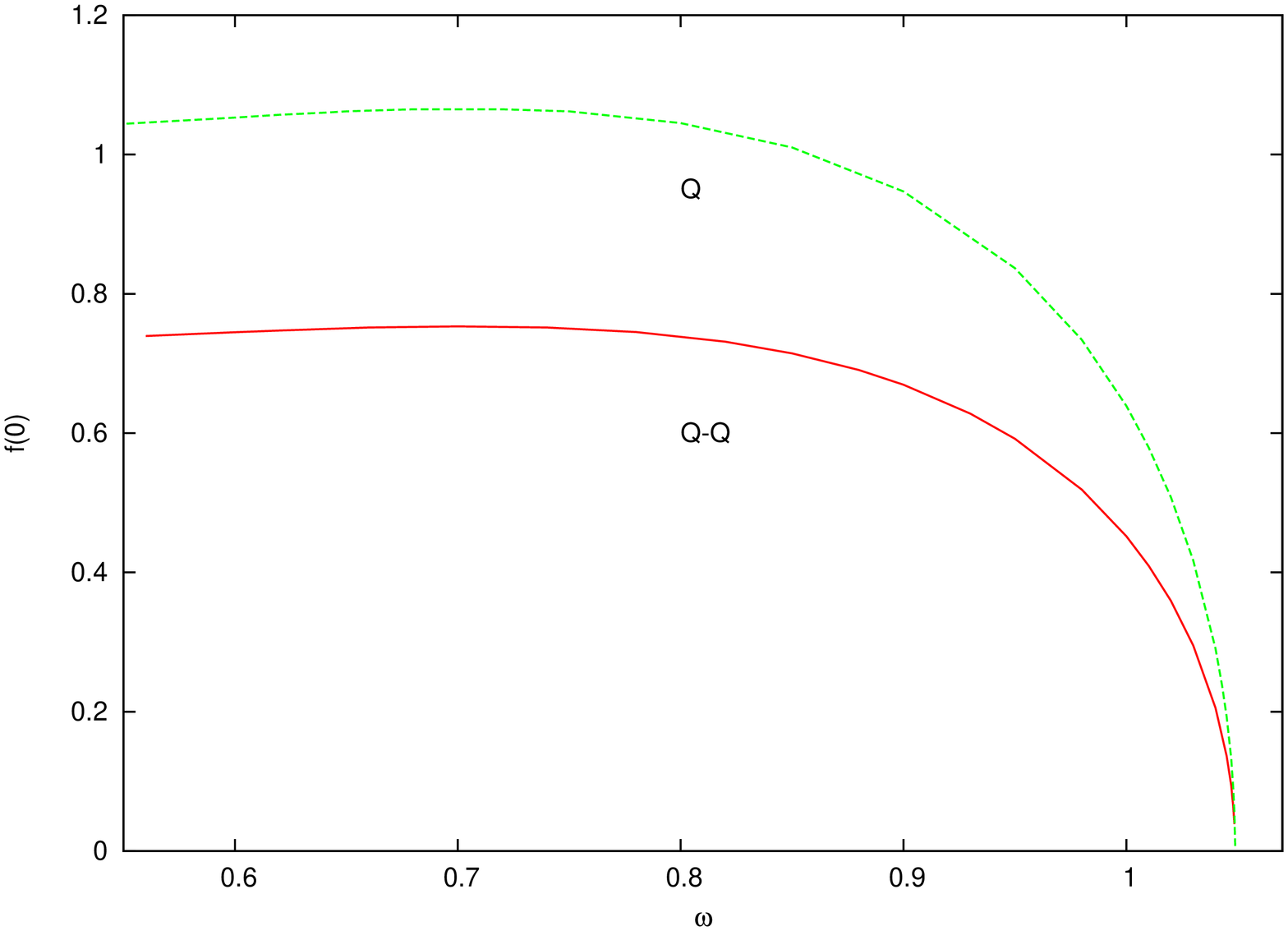}
\end{center}
\vspace{-0.5cm}
\caption{\small Energy-charge ratio versus
$\omega$  is shown
for the coupled spherically symmetric 2-Q-ball system and for the single Q-ball (a).
A straight dashed line $E=mQ$ indicating the margin of stability is drawn.
The value of the profile functions $f(0)$ at the origin is shown as function of $\omega$  (b).
}
\end{figure}

\begin{figure}[hbt]
\lbfig{fig:8}
\begin{center}
(a)\hspace{-0.8cm}
\vspace{-0.8cm}
\includegraphics[height=.30\textheight, angle =-90]{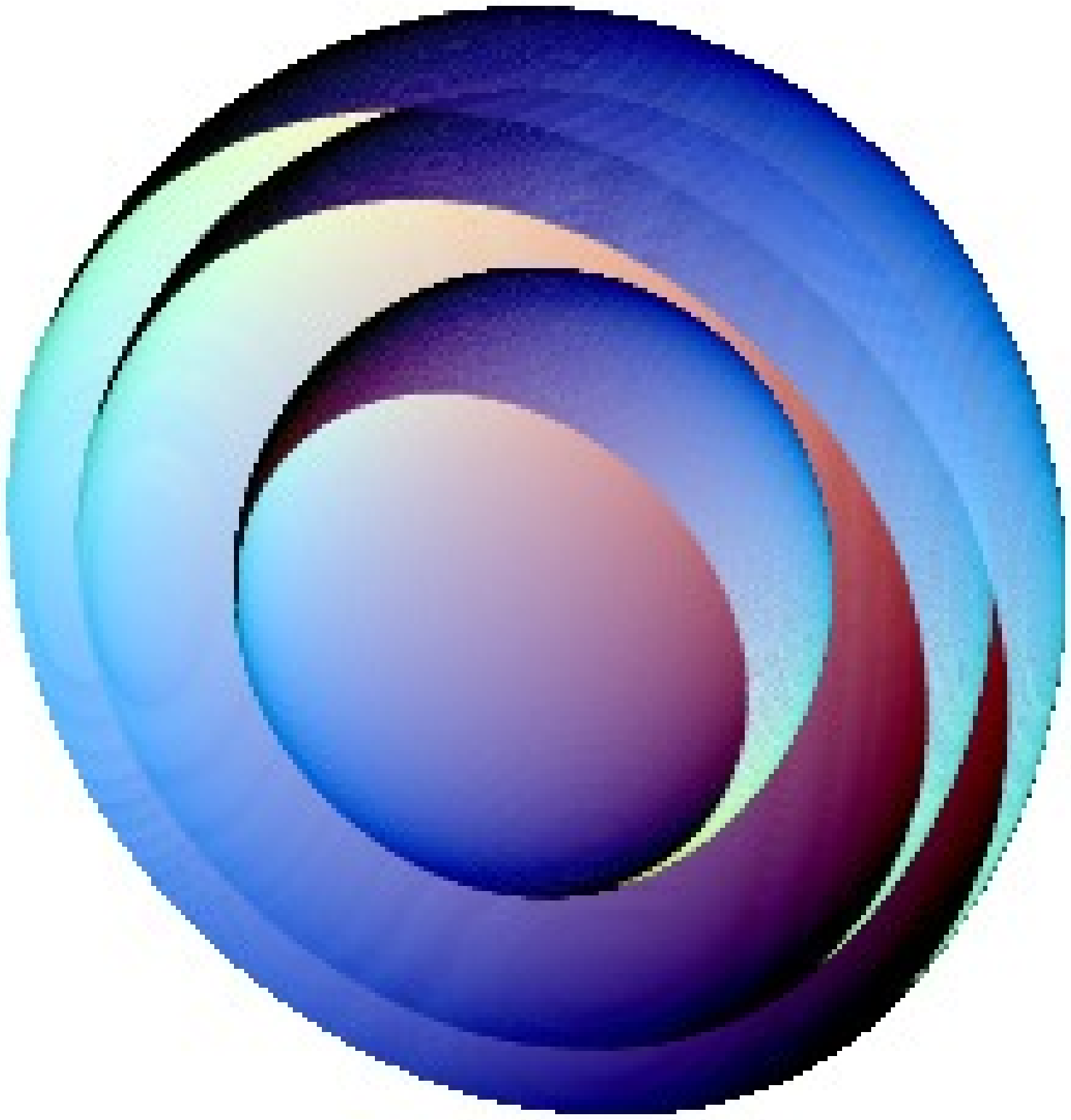}
\hspace{0.5cm} (b)\hspace{-0.6cm}
\includegraphics[height=.36\textheight, angle =-90]{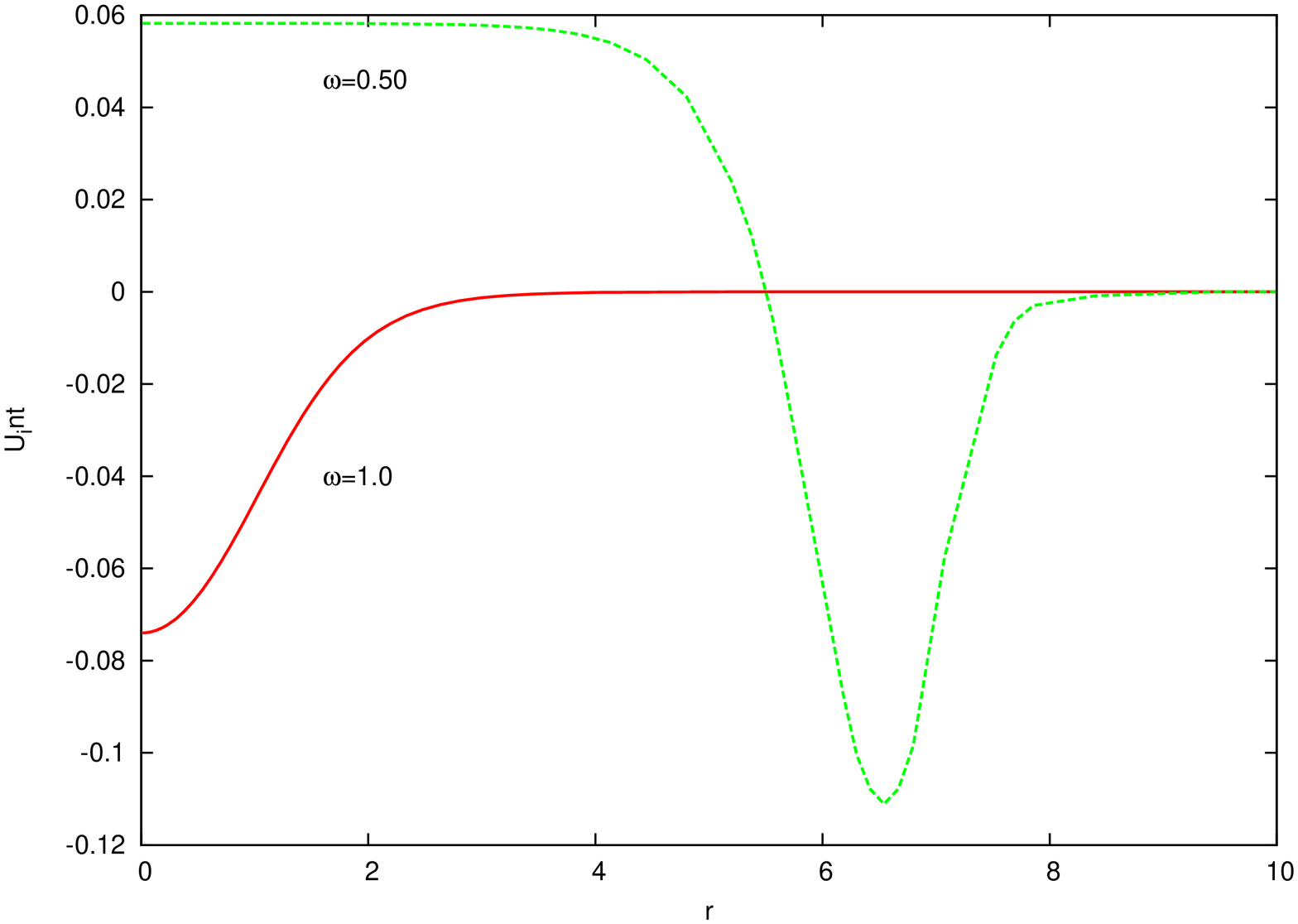}
\end{center}
\vspace{0.5cm}

\caption{\small The energy isosurface of the 2-Q-ball system is shown for
$\omega=0.55$ (a). The potential of interaction is plotted as function of $r$ for
$\omega=1.04$ and $\omega=0.50$ (b).
}
\end{figure}

For two axially symmetric non-spinning coupled Q-vortices and Q-walls we observe a similar pattern. An initial configuration
which satisfies the proper boundary conditions, rapidly converges to the two-component solution
with $X(\rho) = Y(\rho)$, however
the energy and the charge per unit length (area) of the coupled system almost coincide with corresponding values of
the energy and the charge of the single-component fundamental solution over all range of values of $\omega$
presented in Figs.~\ref{fig:3},~\ref{fig:6}. Evidently, the energy/charge relation on the upper branch
is rather different from that of the solitons in the Faddeev-Skyrme model.
Thus, the energy and the charge are equally distributed between the components and the values of the profile functions on
the symmetry axis (symmetry plane) is smaller than they are in the case of the single-component Q-vortex/Q-wall
(see Fig.~\ref{fig:9} (a,b))

\begin{figure}[hbt]
\lbfig{fig:9}
\begin{center}
(a)\hspace{-0.6cm}
\includegraphics[height=.33\textheight, angle =-90]{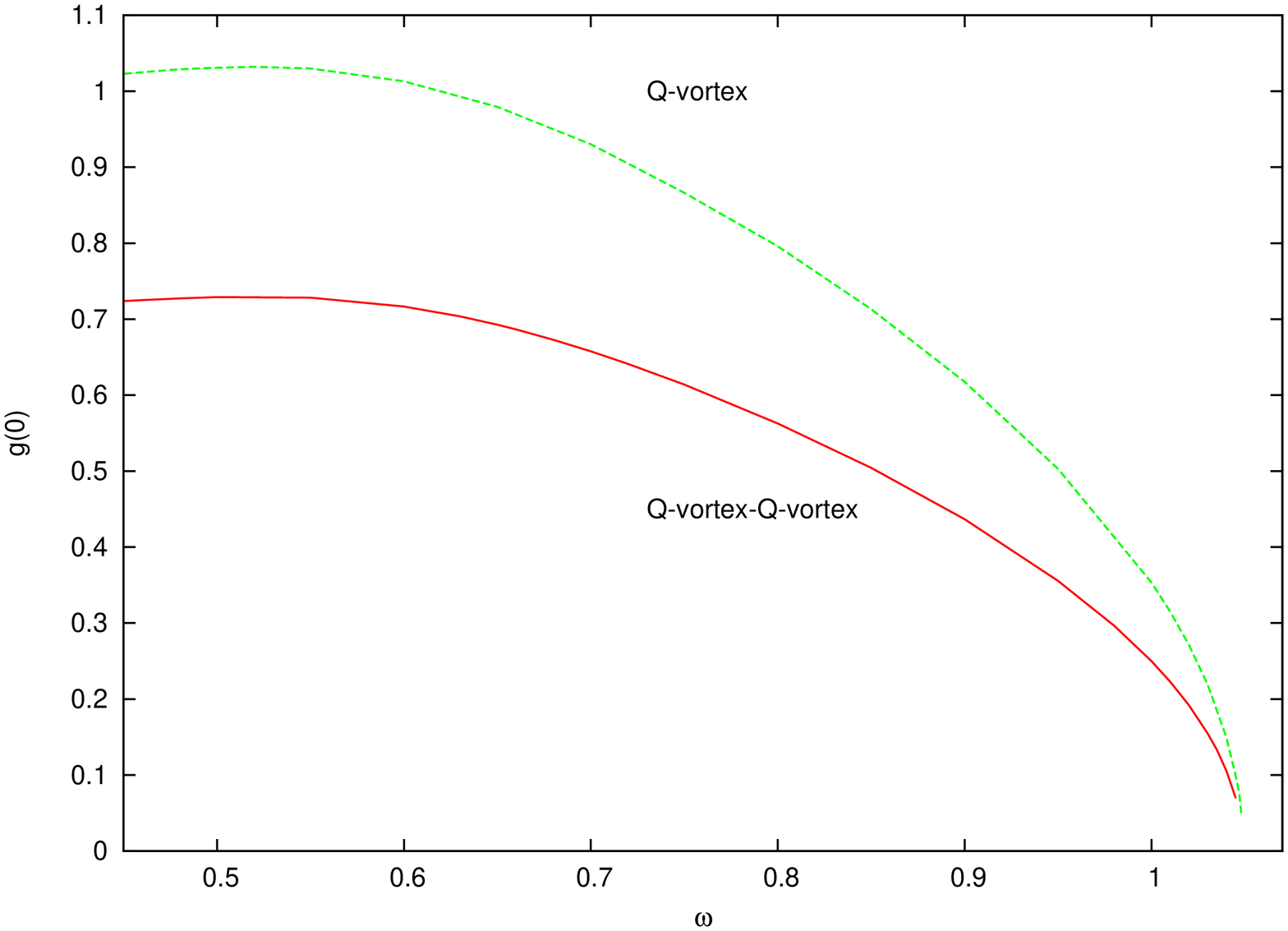}
\hspace{0.5cm} (b)\hspace{-0.6cm}
\includegraphics[height=.33\textheight, angle =-90]{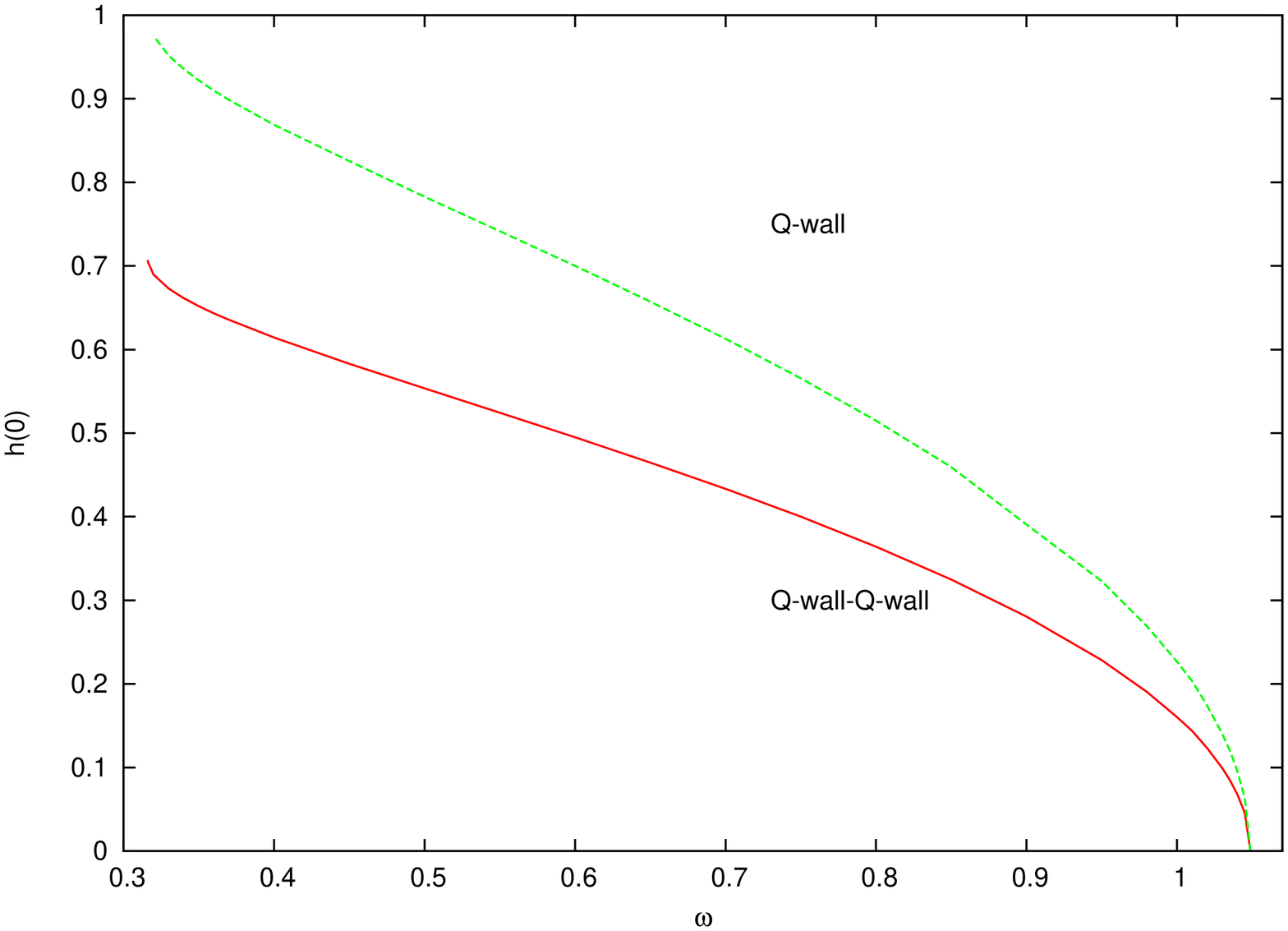}
\end{center}
\vspace{-0.5cm}
\caption{\small 2-Q-vortices (a) and 2-Q-wall (b) coupled system:
The value of the profile functions $g(0)$ and $h(0)$ at the origin are shown as function of $\omega$ together with the
corresponding function of the single component model.
}
\end{figure}

\begin{figure}[hbt]
\lbfig{fig:10}
\begin{center}
(a)\hspace{-0.6cm}
\includegraphics[height=.24\textheight, angle =0]{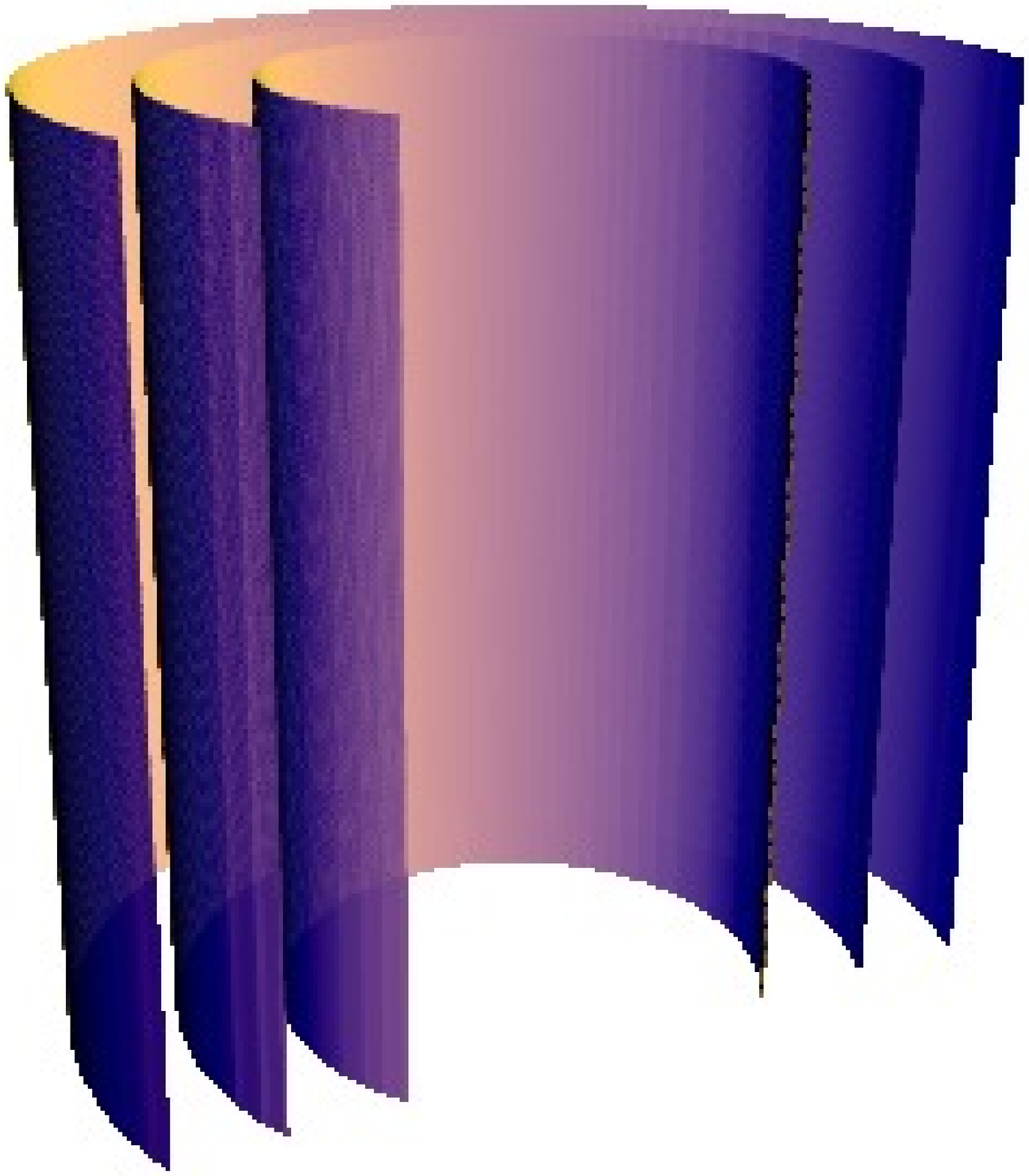}
\hspace{0.5cm} (b)\hspace{-0.6cm}
\includegraphics[height=.24\textheight, angle =0]{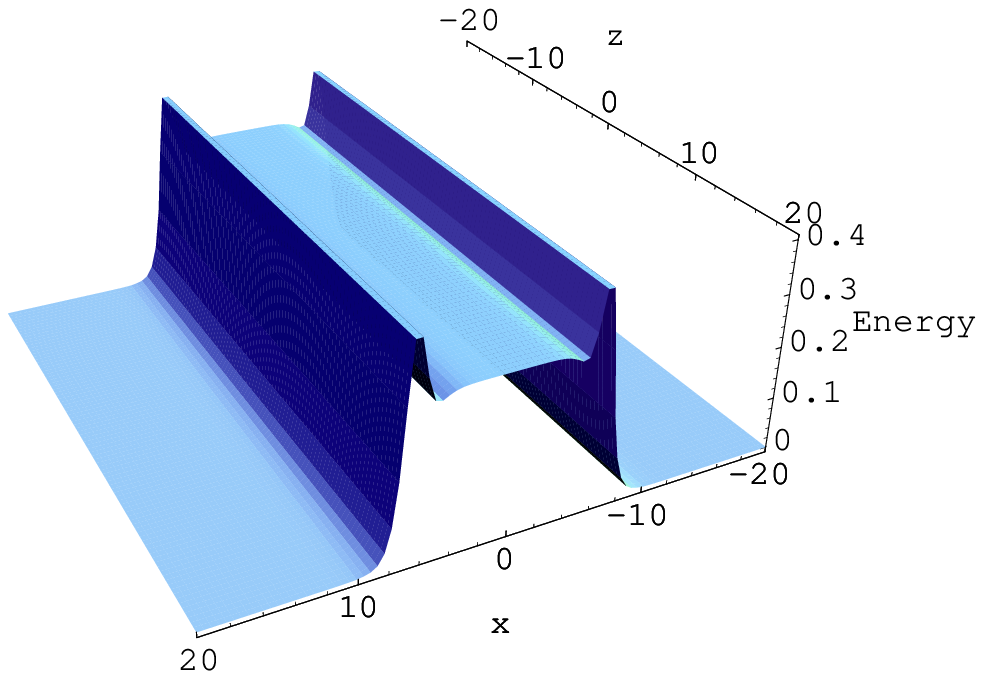}
\end{center}
\vspace{-0.5cm}
\caption{\small
The energy isosurface of the 2-Q-vortex system at $T_0^0 = 0.21$
(a) and the energy density of the 2-Q-vortex configuration
as function of $x,z$ coordinates  (b)
are shown for $\omega = 0.43$.}
\end{figure}

Again, as $\omega \to \omega_+$ and $\omega \to \omega_-$ the behavior of the coupled axially symmetric 2-Q-vortex system and
2-Q-wall configuration
follow the familiar pattern of the single Q-vortex/Q-wall respectively.
As $\omega$ decreases, both components of the coupled system expand,
the potential of interaction then is repulsive on the symmetry axis/symmetry plane and it has a minimum at $\rho = \rho_0(\omega)$ and
$z = z_0$, respectively. Thus the energy density
distribution has the shell structure (see Fig.~\ref{fig:10} (a)). Then both components rapidly expand forming a
new vacuum around the $z$-axis/$x-y$ plane, as seen from Fig.~\ref{fig:10} (b)
(cf similar behavior of the Q-wall, Figs.~\ref{fig:4}, \ref{fig:5}).
Due to severe numerical difficulties encountered here, we could not clarify the properties of the critical solution as
$\omega$ approaches $\omega_-$, in particular it is unclear whether
the bounded 2-Q-vortex system remains stable in this region or it
decays into Q-balls with lower energy per unit length.

\subsection{Coupled spinning $n=1$ solitons}
The ansatz \re{spinning-twisted} also describes spinning configurations with $n\ne 0$. Indeed, it was demonstrated
in \cite{Volkov:2002aj} that the single component
model \re{model} admits  spinning, axially symmetric generalizations for the spherically symmetric Q-balls.
The properties of these solutions were analysed in \cite{Brihaye:2007tn,KKL1,Kleihaus:2007vk,Axenides:2001pi}.
The energy density of these spinning Q-balls is of toroidal shape and they possess the angular momentum
$J = n Q$. It has also been shown by Volkov and W\"ohnert \cite{Volkov:2002aj}
that for a given $n$ both parity-even (i.e. symmetric, $P = +1$) and parity odd (i.e., antisymmetric,
$P = -1$) solutions exist. Furthermore, there are radial and angular unstable excitations
of the fundamental Q-balls which are related to the spherical harmonics \cite{Volkov:2002aj,Brihaye:2007tn}.
The radial excitations are parametrised by the number of nodes of the scalar field $k\in \mathbb{Z}$.
Also the spinning excited Q-vortex solutions are known \cite{Volkov:2002aj}.

However, we observe that no spinning Q-wall configuration with $n\ne 0$
is likely to exist. This agrees with the physical intuition based on a
symmetry argument. However we have found some evidence that there are families of generalized
Q-wall solutions spinning around the $\rho$ axis. This study will be reported elsewhere.

A general property of the spinning 2-soliton solutions parametrised by the ansatz \re{spinning-twisted} for the system
\re{model} is that the interaction between the components rapidly drives the system to the degenerated state
$X(\rho, z)=Y(\rho,z)$. We restrict our discussion to two particular cases considering first
the system of parity-even $X$-component in the sector with $n=1,k=0$ and secondly, the
parity-even angularly excited $Y$-component with $n=1,k=0$. This system
would provide an example of a "non-twisted" (1,0) Q-ball \re{spinning-twisted} with
winding number $m=0$. Similarly, one might try to find a "twisted" Q-ball solution,
employing the parity-even $X$-component in the sector with $n=1,k=1$ and the parity-odd angularly-excited
$Y$-component \cite{Radu:2008pp}. These configurations serve as a first guess to obtain solutions of
the Euler–-Lagrange equations.

The numerical calculations clearly show that  it is energetically favorable in both of these cases to have two component
Q-balls sitting on top of each other with both constituents having identical geometry. Thus, the system converges to
the two copies of the rescaled axially symmetric angularly-radially excited Q-ball which are parity-even and asymmetric,
respectively (see Fig.~\ref{fig:11}).

\begin{figure}[hbt]
\lbfig{fig:11}
\begin{center}
(a)\hspace{-0.6cm}
\includegraphics[height=.34\textheight, angle =-90]{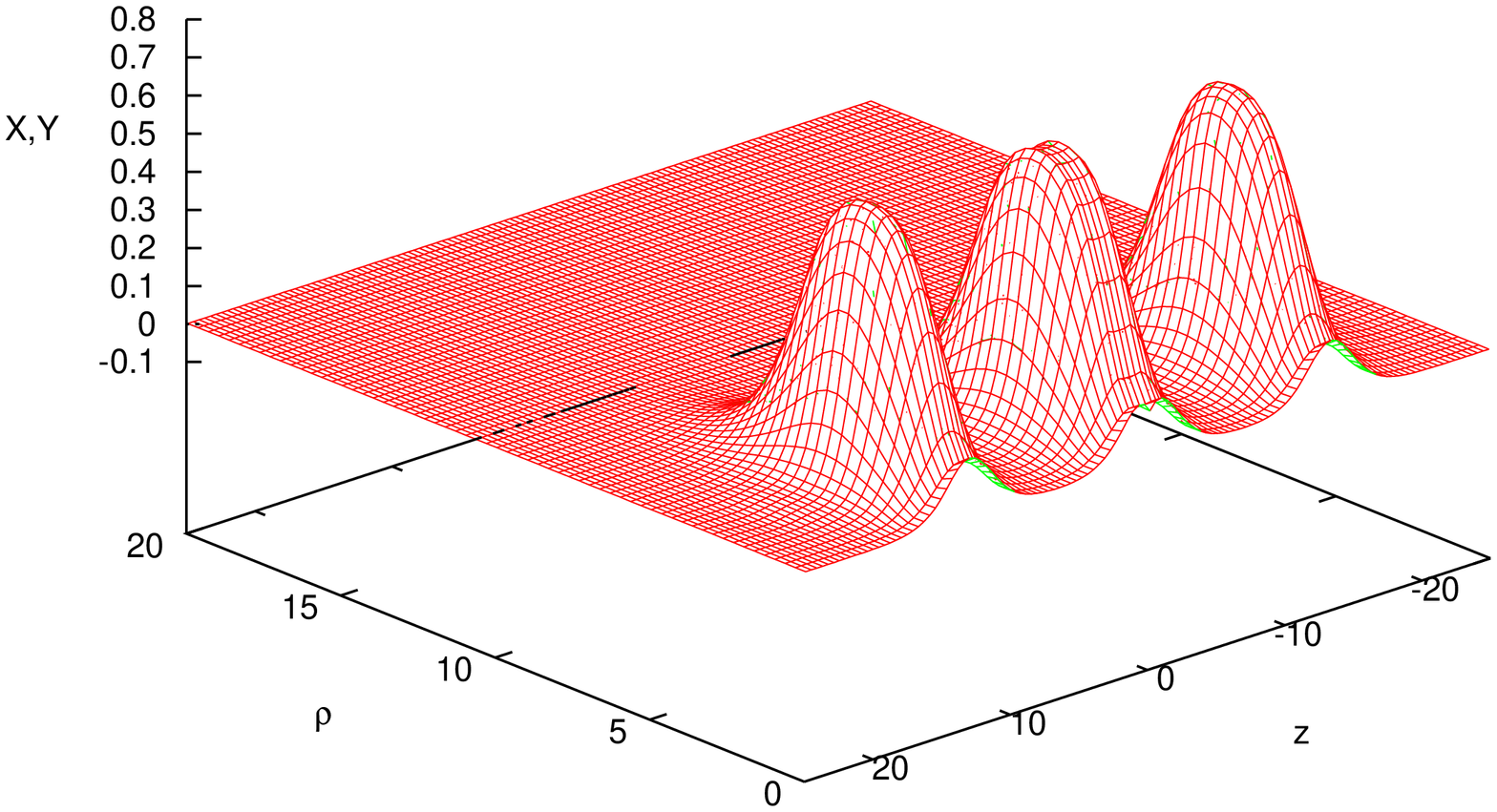}
\hspace{0.5cm} (b)\hspace{-0.6cm}
\includegraphics[height=.34\textheight, angle =-90]{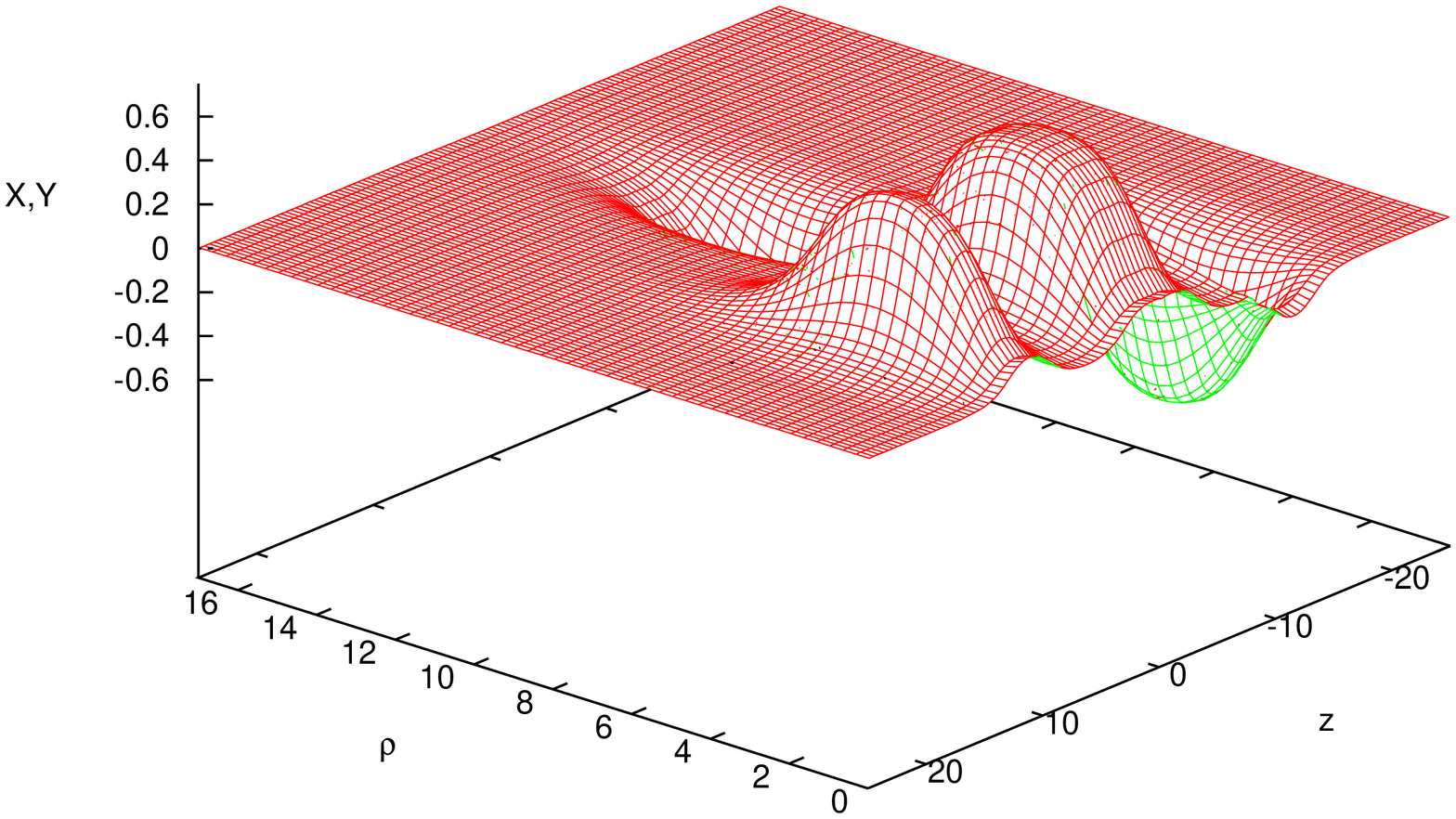}
\end{center}
\vspace{-0.5cm}
\caption{\small
The profile of the function $X(\rho,z)=Y(\rho,z)$ is shown for the parity even
"non-twisted" (1,0) Q-ball (a) and for the mixed parity "twisted" (1,1) Q-ball (b)
for $\omega = 0.70$.}
\end{figure}


\begin{figure}[tbh]
\begin{center}
\setlength{\unitlength}{1cm}
\lbfig{fig:12}
\begin{picture}(0,9.0)
\put(-6.4,0.5)
{\mbox{
\psfig{figure=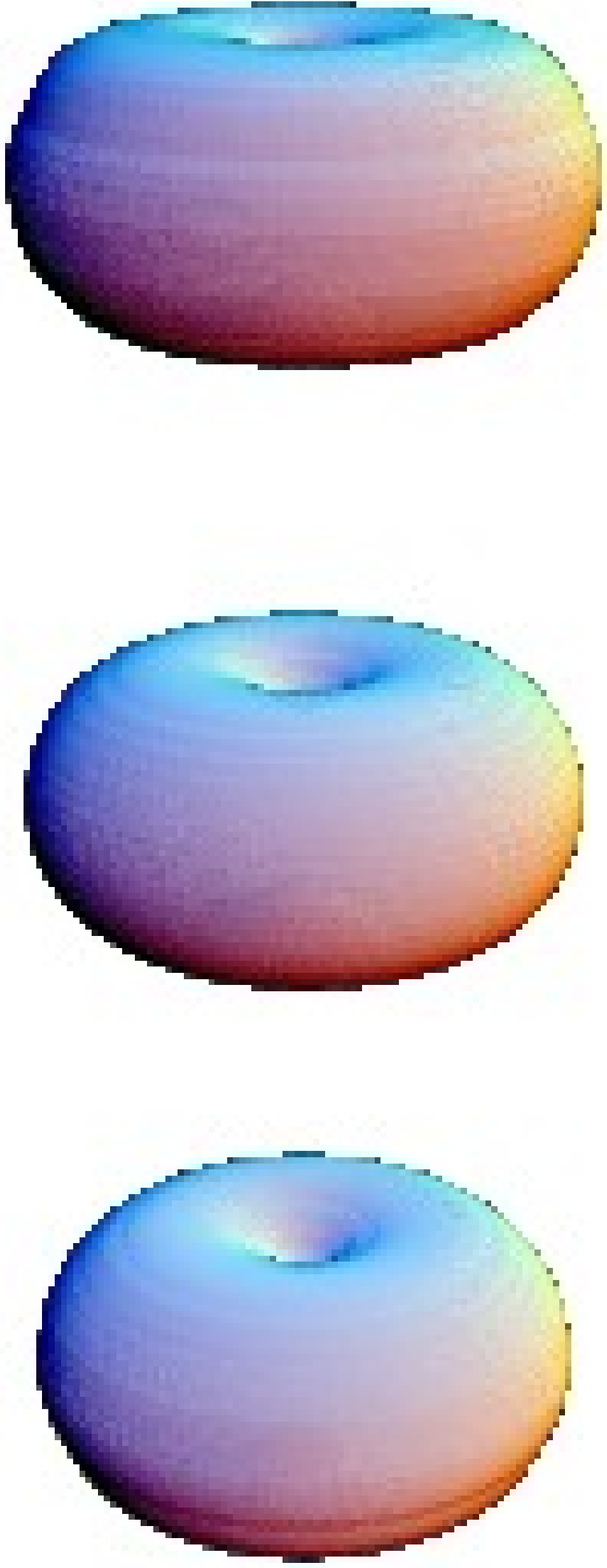,height=6.7cm, angle =0}}}
\end{picture}
\begin{picture}(0,0.0)
\put(0.0,-0.8)
{\mbox{
\psfig{figure=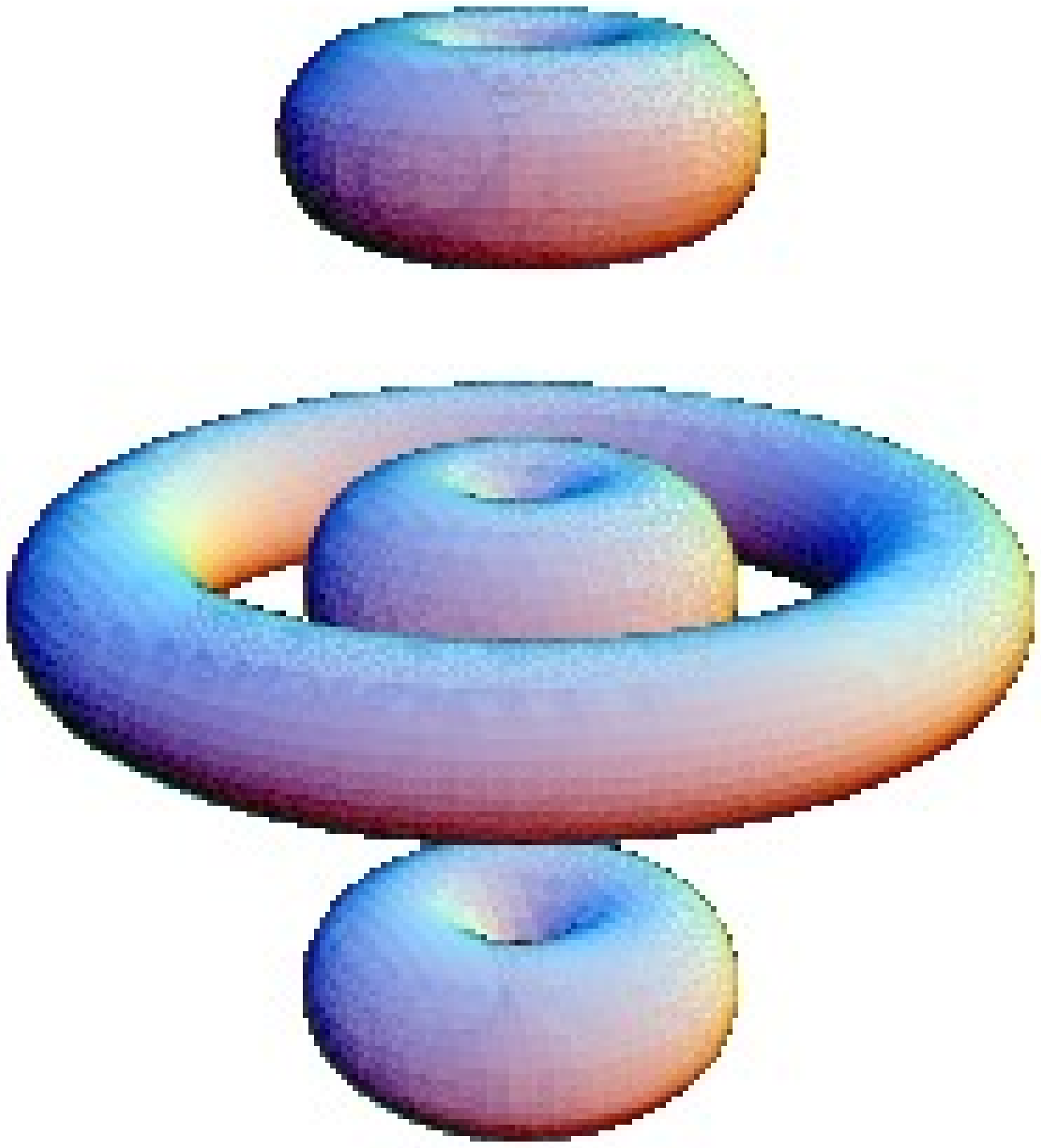,height=7.9cm, angle =0}}}
\end{picture}
\end{center}
\caption{\small
Isosurfaces of constant energy density of the "non-twisted" (1,0) Q-ball with
$T_0^0 = 0.32$ (left frame) and the "twisted" (1,1) Q-ball with $T_0^0 = 0.35$ (right frame)
are shown for $\omega = 0.90$.}

\end{figure}

As seen in Fig.~\ref{fig:11} (a) for the parity even axially symmetric spinning (1,0)  2-Q-ball configuration
the fields have three almost identical maxima at a finite distance from the $z$-axis with
one maximum in the equatorial plane and two other maxima located symmetrically with respect to the $x-y$ plane.
The energy-momentum and charge densities then exhibit a system of three tori presented in Fig.~\ref{fig:12},
left frame.

For the asymmetric spinning (1,1) 2-Q-ball configuration the functions $X(\rho,z)=Y(\rho,z)$ have a bit more
complicated stricture which can be interpreted as superimposed field of two excited Q-balls. For the first radially
excited spinning symmetric Q-ball, the amplitude has one node in the
equatorial plane and the energy and charge densities form a system of two concentric tori, for the second
angularly excited antisymmetric spinning Q-ball, the amplitude also has a node in the equatorial plane
and the energy and charge densities show two maxima located symmetrically with respect to the $x-y$ plane as shown
in Fig.~\ref{fig:12}, right frame.

In Fig.~\ref{fig:13} we plotted the energy $E$ of the (1,0) and (1,1) composite 2-Q-ball configurations
as a function of $Q$ (frame (a)) and the ratio $E/Q$ versus $\omega$ (frame (b)).
As before, the straight line $E=mQ$ indicates the region of stability of the configurations with respect
to decays into quanta of the scalar field. In both cases the behavior of the 2-Q-ball systems is almost identical
to what we observed for the single axially symmetric angularly-radially excited Q-balls.
Thus the (1,0) 2-Q-ball becomes unstable as
$\omega$ approaches the critical value $\omega_{cr}^{(1)} \approx 0.835$ and the
(1,1) 2-Q-ball is unstable with respect to decays as
$\omega > \omega_{cr}^{(1)} \approx 0.847$. Both values are smaller than the critical frequency of the
fundamental spherically symmetric Q-ball.

\begin{figure}[hbt]
\lbfig{fig:13}
\begin{center}
(a)\hspace{-0.6cm}
\includegraphics[height=.33\textheight, angle =-90]{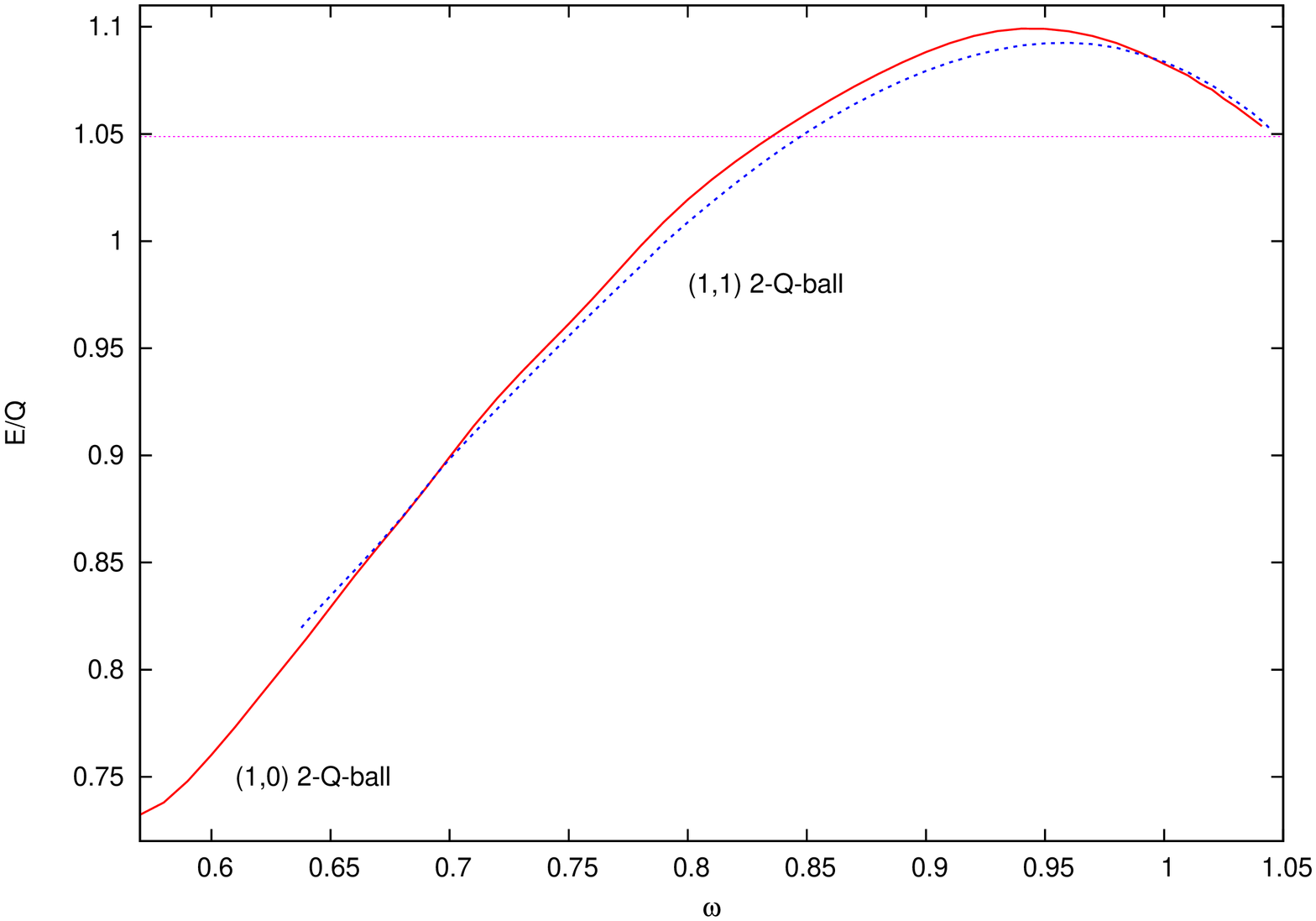}
\hspace{0.5cm} (b)\hspace{-0.6cm}
\includegraphics[height=.33\textheight, angle =-90]{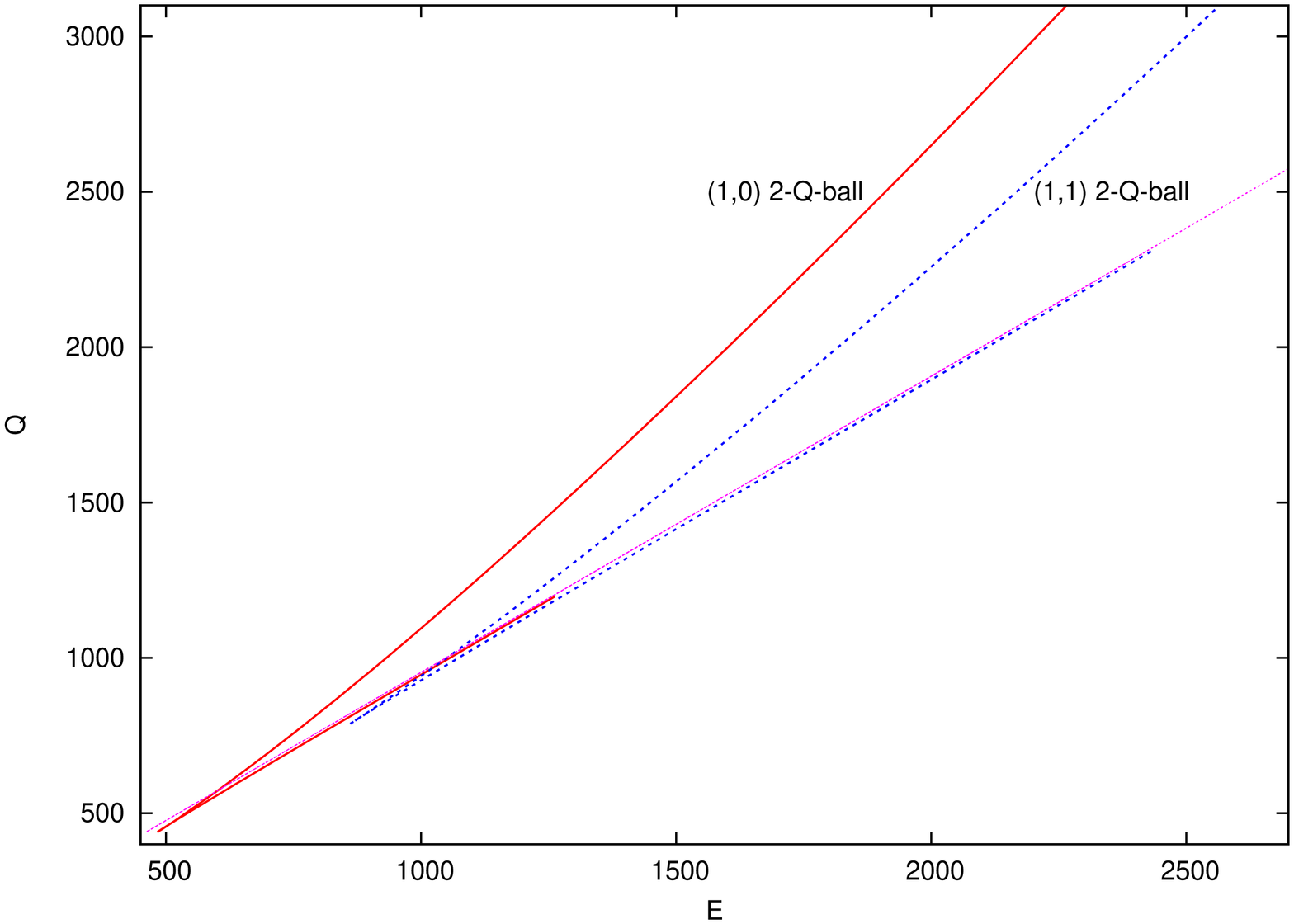}
\end{center}
\vspace{-0.5cm}
\caption{\small Energy-charge ratio versus
$\omega$ (a) and the energy versus charge (b) are shown
for the (1,0) and (1,1) 2-Q-balls. A straight dashed line $E=mQ$ indicating the margin of stability is drawn.
}
\end{figure}

Charge and energy then approach a minimum forming a typical spike profile
(see Fig.\ref{fig:13}, frame (b)) which indicates the saddle node bifurcations and
instability of the upper branch. On the other hand, the composite
spinning 2-Q-balls parametrized by the ansatz \re{spinning-twisted}
are not stable with respect to decay into the fundamental spinning Q-balls. Numerical calculations reveal
this instability as $\omega$ decreases below a certain threshold value which is much higher than the
$\omega_-^2 = 0.1$, for example the (1,0) and (1,1) 2-Q-balls are unstable with respect to decay into
the coupled system of 2 radially excited $n=1$ Q-balls as $\omega$ approaches the second critical value
$\omega_{cr}^{(2)} \sim 0.60$.

Let us finally briefly discuss one more possibility
to construct a composite system of two coupled non-topological solitons
\cite{Brihaye:2007tn,Radu:2008pp}.
We follow closely the approach described in \cite{Brihaye:2007tn}.
The Lagrangian density for this model is given by
\be \label{model-yves}
L = L(\Phi_1) + L(\Phi_2) - \lambda |\Phi_1|^2 |\Phi_2|^2
\ee
where two copies of the Lagrangian \re{model}
$L(\Phi_i)=\partial_\mu \Phi_i \partial^\mu \Phi_i^* - U(|\Phi_i|)$, $i=1,2$ are coupled through
the minimal interaction term.
This coupling is much less restrictive than the example \re{spinning-twisted}  above. Indeed, both the coupling constant
$\lambda$, the frequencies $\omega_1, \omega_2$ and the windings $n_1,n_2$ are free parameters of the model.
As compared to the parametrization \re{spinning-twisted}, we expect
the existence of a much richer set of possible composite solutions for the model \re{model-yves}.
Some of them were considered in
\cite{Brihaye:2007tn,Radu:2008pp}, here we would like to note that it is now possible, within this model, to couple
two solitons having different geometry. In Fig.~\ref{fig:14} we exhibit the energy energy isosurfaces of some
two-component configurations at $\lambda=0.1$ and $\omega_1=\omega_2=0.90$:
\begin{itemize}
    \item[(a)] Q-wall ( $n_1=0$) and the fundamental Q-ball ($n_2=0$);
    \item[(b)] Q-vortex ($n_1=0$) and Q-wall ($n_2=0$);
    \item[(c)] Q-vortex ($n_1=0$) and spinning Q-ball ($n_2=1$) (`hoop' solution \cite{Radu:2008pp});
    \item[(d)] Q-wall ($n_1=0$)  and spinning Q-ball ($n_2=2$);
    \item[(e)] Q-vortex ($n_1=0$) and the even parity angularly excited spinning Q-ball ($n_2=2$)(double hoop configuration);
    \item[(f)] Q-wall ($n_1=0$) and the odd parity angularly excited spinning Q-ball ($n_2=1$).
\end{itemize}
\begin{figure}[tbh]
\begin{center}
\setlength{\unitlength}{1cm}
\lbfig{fig:14}
\hspace{0.5cm} (a) \hspace{6.0cm} (b) \hspace{7.0cm} (c) \\
\begin{picture}(0,8.0)
\put(-0.5,2.1)
{\mbox{
\psfig{figure=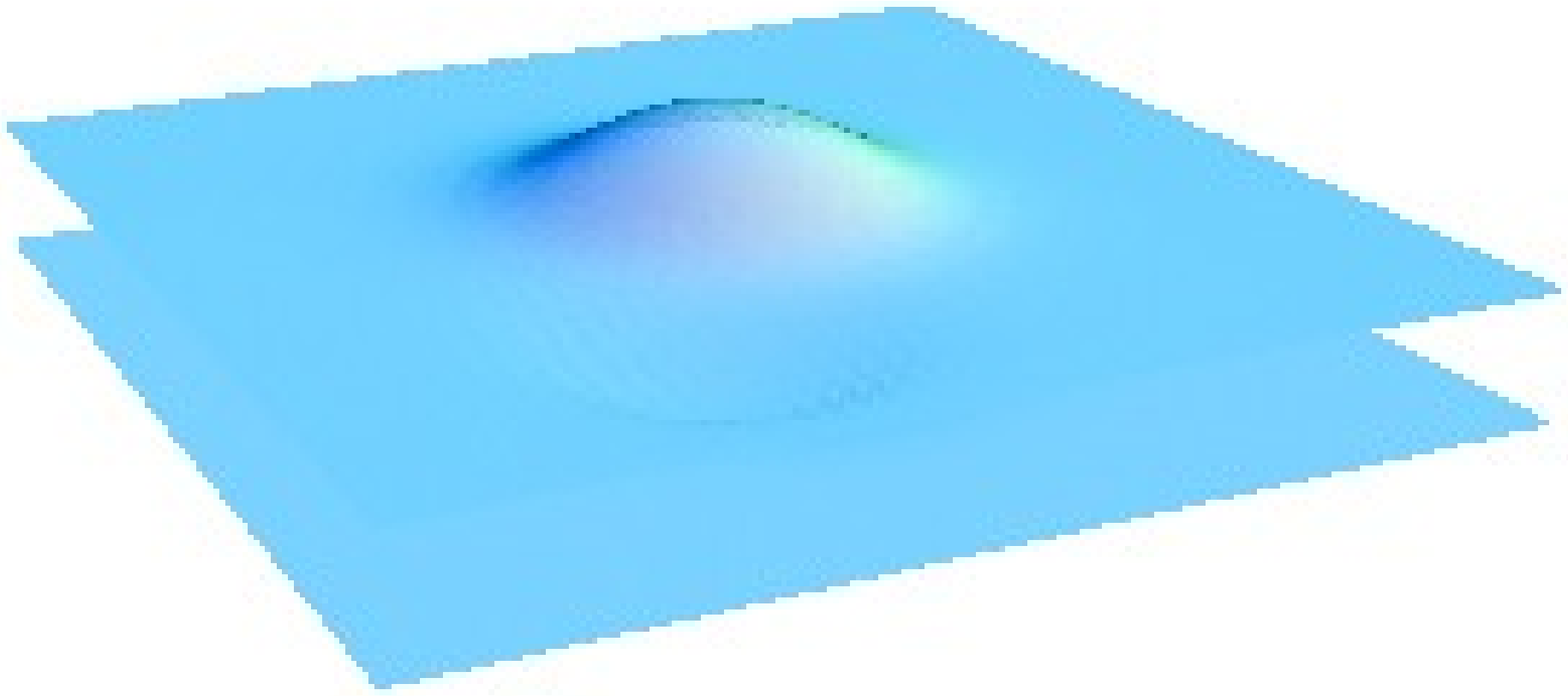,height=3.2cm, angle =0}}}
\end{picture}
\begin{picture}(0,1.7)
\put(5.1,0.5)
{\mbox{
\psfig{figure=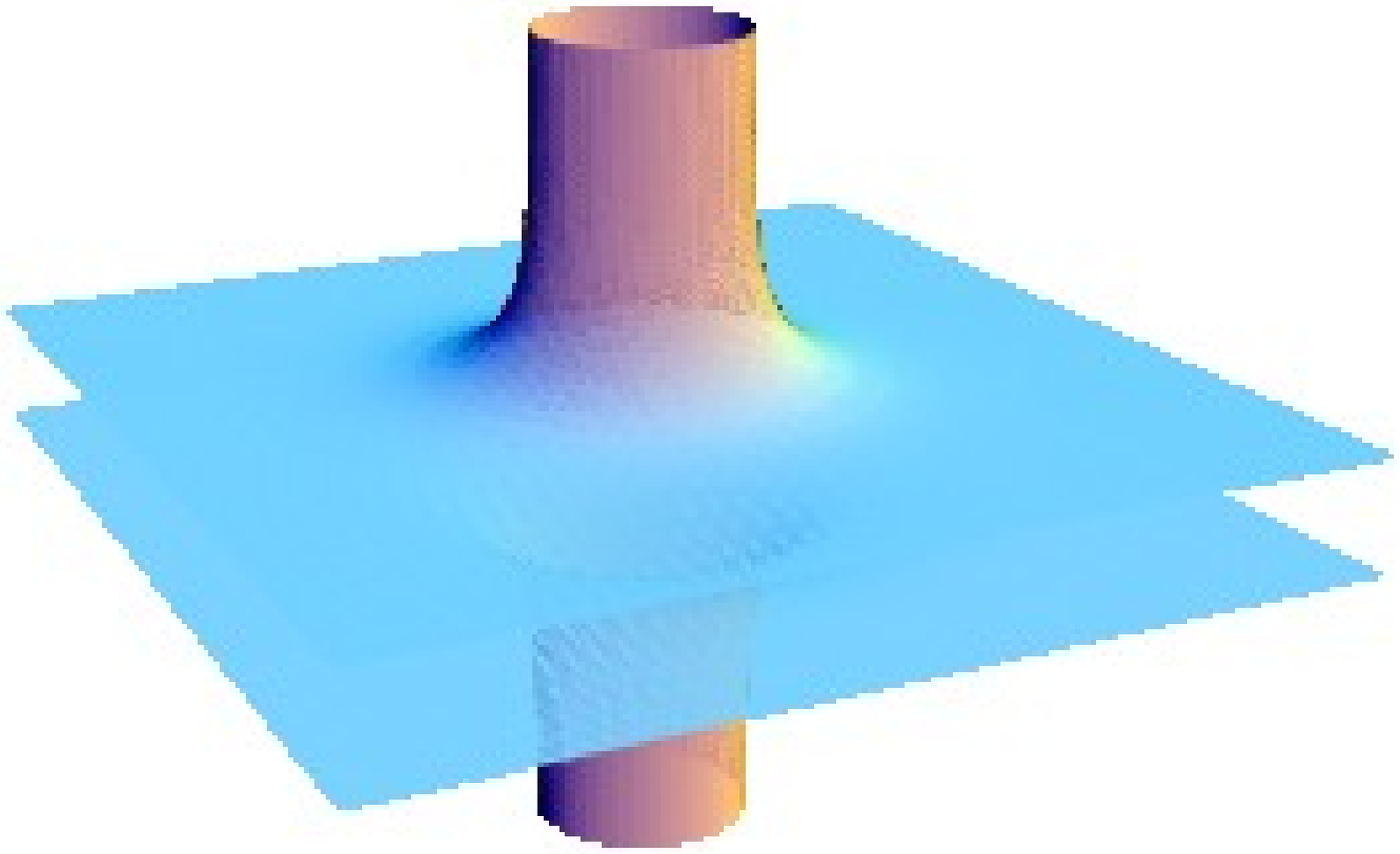,height=5.9cm, angle =0}}}
\end{picture}
\begin{picture}(0,0.0)
\put(12.0,0.8)
{\mbox{
\psfig{figure=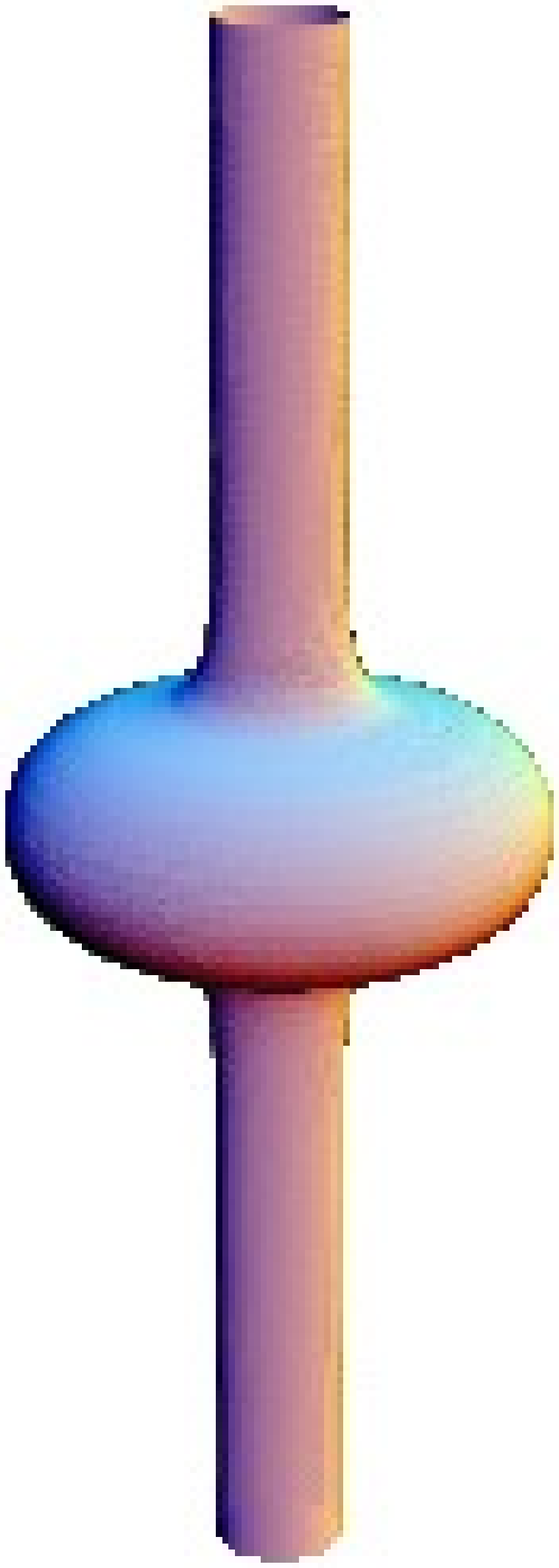,height=6.2cm, angle =0}}}
\end{picture}
\hspace{0.5cm} (d) \hspace{6.0cm} (e) \hspace{7.0cm} (f) \\
\begin{picture}(0,6.2)
\put(-8.0,0.5)
{\mbox{
\psfig{figure=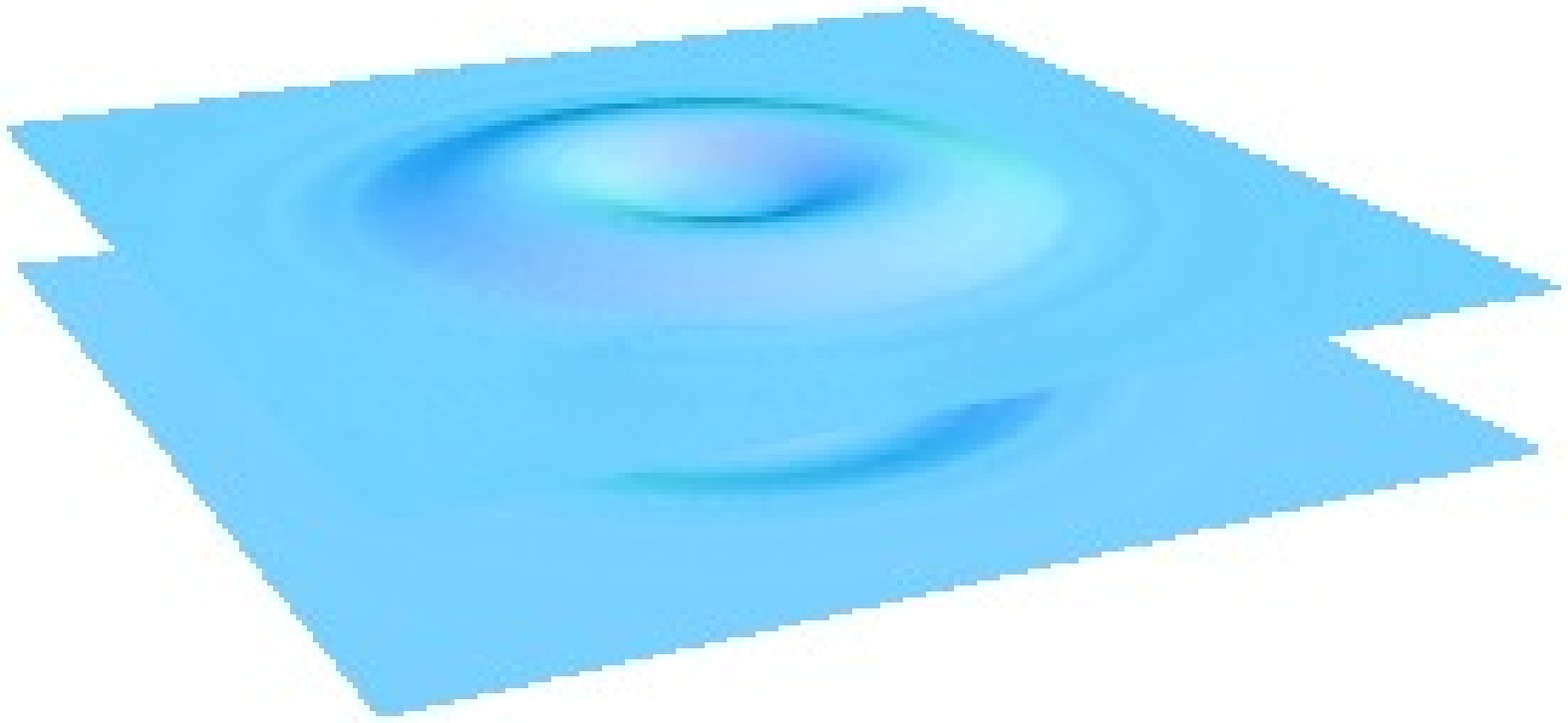,height=3.1cm, angle =0}}}
\end{picture}
\begin{picture}(0,1.7)
\put(-1.4,-0.5)
{\mbox{
\psfig{figure=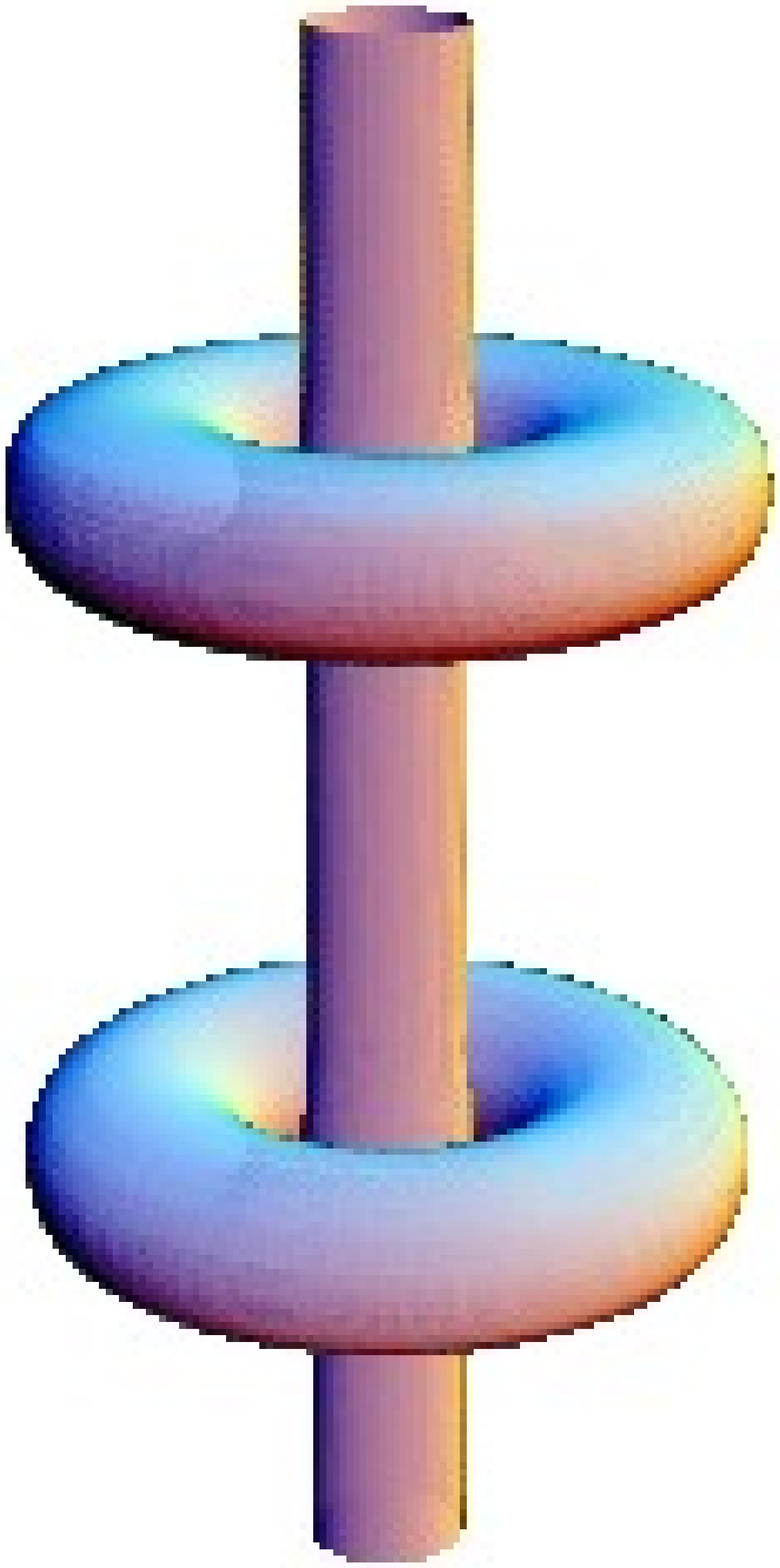,height=5.9cm, angle =0}}}
\end{picture}
\begin{picture}(0,0.0)
\put(2.2,-1.0)
{\mbox{
\psfig{figure=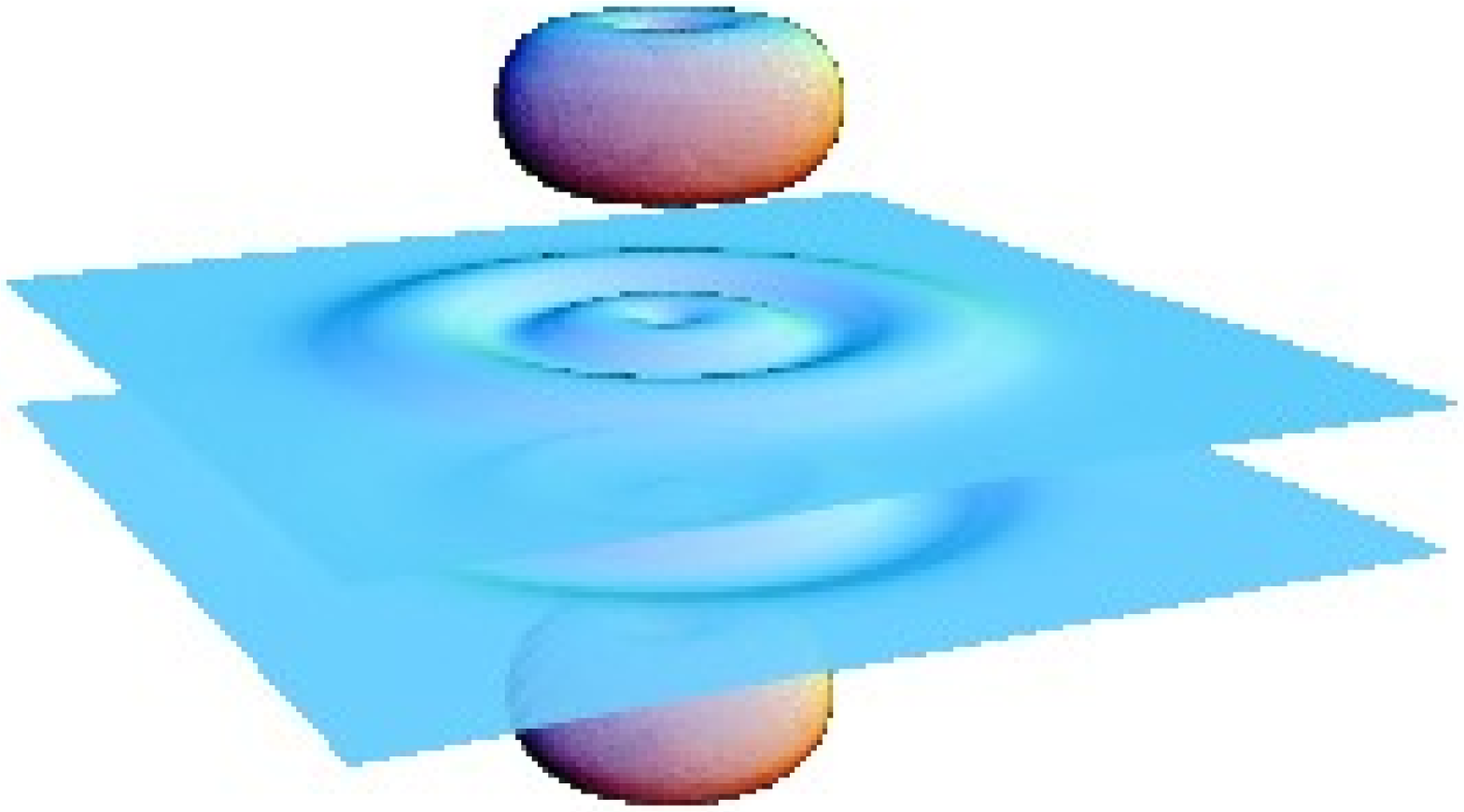,height=5.2cm, angle =0}}}
\end{picture}
\end{center}
\caption{\small The energy isosurfaces of the coupled system with component fields of different geometry are shown for
$\omega_1=\omega_2=0.90$ and coupling $\lambda = 0.1$.
}
\end{figure}

Obviously the stability of these configurations depends on the coupling $\lambda$.
The investigation of these and other two-component solutions of the model
\re{model-yves} and their eventual bifurcations is currently underway.

\section{Conclusions}
Motivated by the recent interest in non-topological stationary solitons
possessing various types of symmetry, we investigated properties of the
corresponding solutions to the same simple model with sextic potential:
configurations with spherical symmetry (Q-balls),  cylindrical symmetry (Q-vortices)
and planar symmetry (Q-walls).

The numerical studies of these three types of fundamental solutions show different behavior of
charge and energy with changing $\omega$ for each type.
Their dependence on  $\omega$ is
analysed numerically in some detail. By and large this is qualitatively very similar to
that for the spherically symmetric solutions~\cite{Coleman:1985ki}.
In all cases there is a different behavior as $\omega$ tends to $\omega_+$ or  $\omega \to  \omega_-$.
We have also found numerical evidence of an instability of the Q-vortices and the
Q-walls in the limit where $\omega \to  \omega_-$.
Our numerical investigations indicate fairly clearly that there is no spinning generalization of the Q-wall solution, which
would be constructed by a direct analogy with spinning Q-balls and Q-vortices.
It does not exclude another possibility that there are  generalized
Q-wall solutions spinning around an axis in the x-y plane.

Also, we have studied a coupled two-component system using the generalized field ansatz which includes an
independent phase \cite{Radu:2008pp} and have constructed several examples which represent both spinning
and non-spinning configurations. Previous research has suggested that such  a system may support
the existence of `twisted' Q-balls which would have a certain similarity
with twisted loops in the Faddeev-Skyrme model. As it turns out these coupled configurations do not exist
if the constituents possess different geometry.
If the constituents are of the same geometry, however,
then it is energetically favorable to have two component
sitting on top of each other. Thus, the `twisted' system rapidly converges to
the two rescaled copies of the single component model with the energy and the charge
equally distributed between the components.

However such a composite system is unstable not only with respect to radiation of the scalar quanta as
$\omega \to \omega_+$, but also with respect to decay into the fundamental solitons.
Numerical calculations reveal this instability as $\omega$ decreases below certain
threshold value which is much higher than the $\omega_-$.

Another possibility, discussed in \cite{Brihaye:2007tn}, is not so restrictive. A minimal interaction between
two components allows us to construct a plethora of coupled Q-balls with various geometry. The
numerical work involved in the construction of these 2-Q-ball solutions is, however, a considerably challenging task.


\newpage
\bigskip
\noindent
{\bf\large Acknowledgements} \\
This research is inspired by numerous discussions with
Eugen Radu who suggested the general strategy of the research and provided his
related results. I thank Jutta Kunz, Burkhard Kleihaus and Mikhail Volkov
for valuable discussions during various stages of this work. I am indebted
to Derek Harland for a careful reading of the manuscript and his valuable comments.
I gratefully acknowledge support by the Alexander von Humboldt Foundation and
would like to acknowledge the hospitality at the Institut of Physics,
University of Oldenburg.


\end{document}